%% file: temporal-skinning.tex
\documentclass{svproc}
\usepackage{graphicx}
\usepackage{caption}
\usepackage{subcaption}
\usepackage{float}
\usepackage{hyperref}
\usepackage{amsmath}
\usepackage{multirow}
\usepackage{array}
\usepackage{booktabs}
\usepackage{adjustbox}
\usepackage{xcolor}

\usepackage{url}

\begin{document}
\mainmatter              
\title{Temporal Parameter-free Deep Skinning of Animated Meshes}
\titlerunning{Temporal Parameter-free Deep Skinning of Animated Meshes}  
%
\author{Anastasia Moutafidou\inst{1} \and Vasileios Toulatzis\inst{1}
\and Ioannis Fudos\inst{1}}

\authorrunning{A. Moutafidou, V. Toulatzis and I. Fudos} 
%
%
\institute{University of Ioannina, Grecce}

\maketitle              

\begin{abstract}
In computer graphics, animation compression is essential for efficient storage, streaming and reproduction of animated meshes. Previous work has presented efficient techniques for compression by deriving skinning transformations and weights using clustering of vertices based on geometric features of vertices over time. In this work we present a novel approach that assigns vertices to bone-influenced clusters and derives weights using deep learning through a training set that consists of pairs of  vertex trajectories (temporal vertex sequences) and the corresponding weights drawn from fully rigged animated characters. The approximation error of the resulting linear blend skinning scheme is significantly lower than the error of competent previous methods by producing at the same time a minimal number of bones. Furthermore, the optimal set of transformation and vertices is derived in fewer iterations due to the better initial positioning in the multidimensional variable space. Our method requires no parameters to be determined or tuned by the user during the entire process of compressing a mesh animation sequence.
\end{abstract}
\input{Contents/Introduction.tex}
\input{Contents/RelatedWork.tex}
\input{Contents/Method.tex}

\input{Contents/Experiments.tex}
\input{Contents/Conclusions.tex}


\clearpage

%
%
\bibliographystyle{spmpsci}
\bibliography{temporal-skinning} 

\end{document}

%% file: Contents/Introduction.tex
\section{Introduction}
\label{sec:Introduction}


Nowadays an animator may produce a realistic character animation by following either of the two modern workflows:
\begin{itemize}
	\item[i] rigging a static mesh (i.e. define a bone structure and associate the bones with the mesh vertices by weight painting), apply transformations to bones along a time line, correct erroneous deformations by adding bones, introduce additional per frame deformations to simulate non linear effects, or
	\item[ii]  use recent developments of computer vision and tracking techniques to derive mesh sequences that are reconstructed by markerless capture or by motion capture with dense markers (see e.g.\cite{jacobson-2014}).
\end{itemize}

Both workflows produce sequences of animated meshes. These mesh animation sequences must subsequently be converted to a representation that allows for streaming and editing. To this end, a first step is to use compression. 

With the evolution of cloud based graphics applications, a compression approach such as Linear Blend Skinning is a necessity for efficiently storing and using animation sequences. Compression is performed by producing an approximation of the animation that consists of an initial pose and a number of transformations that describe each subsequent pose by a deformation of a surface part.

Linear blend skinning (LBS) \cite{thalmann-1988} is a time and space efficient mesh 
deformation technique where mesh vertices are influenced by a set of bones. 
In spite of several limitations that have been addressed in the literature \cite{jacobson-2014}, 
LBS-based approaches are significant in the animation industry due to their simplicity and straightforward GPU implementation. 

There exists a variety of approaches for compression using clustering techniques, most of which are based on geometric features of vertices over time. 
We introduce a novel deep learning approach that uses  a training set of successfully fully rigged animated models to produce
a skinning model. Given a new animated mesh sequence, the trained network derives pseudo-bones and weights. There is no limit on the number of vertices, faces and frames given as input. There is just an upper limit on the number of bones for all animations. A single trained network can be used to predict weights for any animation sequence.  

We also improve the efficiency of the least square optimization of transformations and weights that is commonly used to reduce the approximation error by employing conjugate gradient optimization that is suitable for multidimensional systems. 

While previous skinning approaches use a predetermined number of bones and several other tuning parameters, our approach is parameter free. An appropriate set of bones is derived 
based on similar successfully rigged animations of the training dataset. 
In our method there is no need for preprocessing (scaling, rotation or translation) for the geometry of the input, since we only use vertex trajectories, so only the relative movement of the vertex is taken into account.
To evaluate our approach, we use both mesh sequences that are derived from rigged and animated characters and benchmark mesh sequences from available animation sequence datasets.
Our experimental evaluation shows that our approach outperforms previous approaches in terms of both compression rate and approximation error for all datasets.

The mesh clustering derived by our method can also be used to create a skeletal rig since it yields segmentations that correspond to the influence of bones on mesh vertices. Therefore, the output of our methods can be easily converted to a fully rigged animated character and used in subsequent phases of animation editing and rendering.


%% file: Contents/RelatedWork.tex
\section{Related Work}
\label{sec:Related Work}

Although there is a lot of research on skinning of animated models, the use of deep learning techniques for skinning has not been explored thoroughly. 

Elastiface	\cite{Zell:2013:EMB:2486042.2486045} indicate that an animated character can be quite complex and that managing and processing needs cumbersome human intervention and a significant amount of computationally intensive tasks. Moreover techniques such as cross-parameterization \cite{Kraevoy:2004:CCR:1186562.1015811} or procedures that can convert an animated character into an animation sequence have high memory and space requirements. So there is a need for different procedures that can produce animated models in a compressed form without being provided with a skeleton or skin specification \cite{10.1145/1186822.1073206}. 

In the context of animation compression, James and Twigg \cite{10.1145/1186822.1073206} were the first to explore the use of LBS to approximately reproduce articulated characters as a function of their bone movement. Extending this work, Kavan et al. \cite{10.1145/1230100.1230109,10.1145/1230100.1230107} presented a dual quaternion skinning scheme that can compute approximations for highly-deformable animations by suggesting that uniformly selected points on the mesh can act as bones. Both of them are enhance their final skinned approximation by exploiting EigenSkin \cite{10.1145/545261.545286}. FESAM \cite{10.1111:j.1467-8659.2009.01602.x} introduced an algorithm that optimizes all of the skinning parameters in an iterative manner. While FESAM offers high-quality reproduction results, is limited to download-and-play scenarios, since they do not use information about topology and the location of proxy joints is occluded once the optimization process kicks in.

\cite{Vasilakis:2016:PPS:2949035.2949049} introduces a pose-to-pose skinning technique that exploits temporal coherence that enables the full spectrum of applications supported by previous approaches in conjunction with a novel pose editing of arbitrary animation poses, which can be smoothly propagated through the subsequent ones generating new deformed mesh sequences.

\cite{Le:2012:SSD:2366145.2366218} approaches a set of example poses by defining a constrained optimization problem for deriving weights and transformations which yields better results in terms of error. In our method we ensure convexity by an additional equation for each vertex and a non negative least square solver. Then we employ linear solvers and update the weights and the bone transformations successively.


On the other hand, there are techniques that can create a plausible skeleton for a mesh model either by exploiting the movement of vertices to perform mesh segmentation \cite{doi:10.1111/j.1467-8659.2008.01136.x}, by exploiting the mesh structure by performing constrained Laplacian smoothing \cite{Au:2008:SEM:1360612.1360643}, or by analyzing the mesh structure of a set of several sparse example poses \cite{Hasler2010}. Recently, \cite{Le2014} presented a method that first produces a large number of plausible clusters, then reconstructs mesh topology by removing bones and finally performs an iterative optimization for joints, weights and bone transformations. Such methods are competent and produce a fully animated rigged object, but usually are semi-automated since their effectiveness and efficiency depends on setting a large number of parameters that are associated with the structure of the mesh or the specifics of kinematics.  

\cite{NeuroSkinning2019} and \cite{RigNet} predicts a set of vertex weights based on the morphology of a static mesh, by previously training with static meshes and their corresponding weights from animated characters. \cite{avatar} presents a method for automatically rigging a static mesh by matching the mesh against a set of morphable models. Our method predicts weights based on the vertex trajectories  by training with the motion of the vertices over time and the corresponding weights from animated characters and is not restricted by the morphology of the static mesh. 

\cite{Bailey2018} tries to capture non linear deformations that are used in conjunction with a linear system and an underlying skeleton by employing a deep learning technique to determine the non linear part. \cite{NNWarp2020} captures better nonlinear deformations by including in the animation pipeline a light weight neural network (NNWarp) that is known for its rich expressivity of nonlinear functions. 


%% file: Contents/Method.tex
\section{Temporal Deep Skinning}
\label{sec:Method}
Skinning is based on the core idea that character skin vertices are deformed based on the motion of skeletal bones. One or more weights are assigned to each vertex that represent the percentage of influence vertices receive from each bone. With this approach we can reproduce an animation sequence based on a reference pose, the vertex weights and a set of transformations for every frame and bone. 

\begin{figure}[!htb]
	\centering
	\includegraphics[width=8cm]{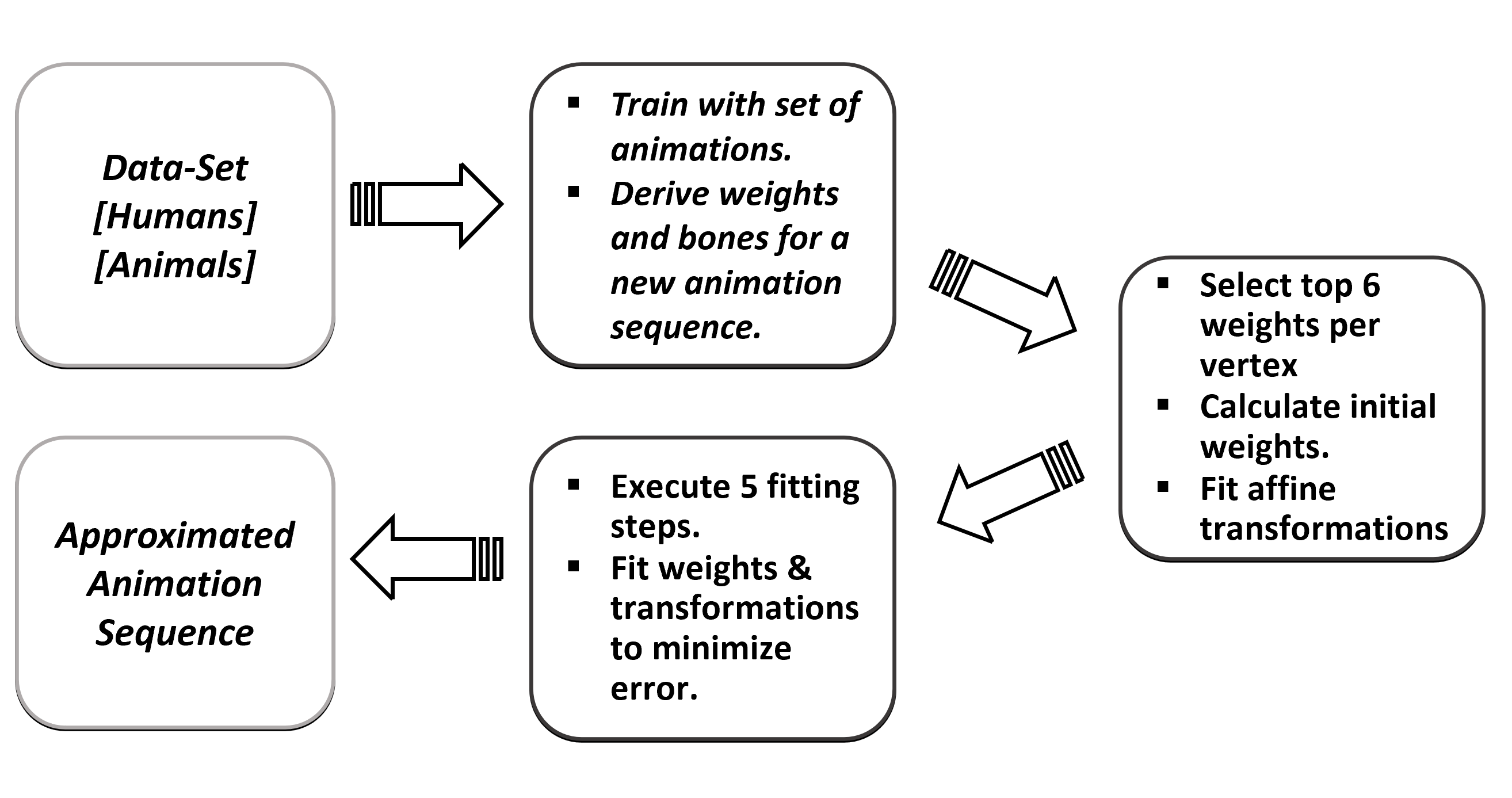}
	\caption{Temporal Deep Skinning workflow.}
	\label{fig:deepworkflow}
\end{figure}

Figure \ref{fig:deepworkflow} illustrates the concept of our {\em Temporal Deep Skinning} method (or simply {\em Deep Skinning} for short). First, we build an appropriate neural network model that classifies each vertex by capturing mesh geometry and vertex kinematics. Then we use a set of human and animal animations to train the neural network model. We achieve this by using as input features the trajectories of all vertices and as output the weights that represent how each vertex is influenced by a bone. The output weight is conceived by the network as the probability of a bone to influence the corresponding vertex. Subsequently, we provide as input to our network arbitrary mesh animation sequences and predict their weights. From the per vertex classifier we determine the number of bones and the weights for each vertex.

We restrict each vertex to have no more than six weights so as to be compatible with the existing animation pipelines \cite{10.1145/1230100.1230109}. For simple gaming characters, usually four weights per vertex are enough, but six weights per vertex can be used to correct artifacts or capture local deformations with pseudo bones. In our comparative evaluation we have implemented all previous methods with six weights as well, so as to conduct an objective comparison. The derived six (or less) weights per vertex correspond to the six highest probability predictions of the network. We then normalize these weights which are already in $[0,1]$ to sum to 1 (coefficients of a convex combination). Since probability prediction of a vertex towards a specific bone cluster represents similarity to a training example, this is naturally translated to influence of the bone on the vertex.

Subsequently, we perform optimization to minimize the least square error between the original and approximated mesh frames. We do so by optimizing weights and transformations in an iterative manner. 

\subsection{Training and Test Datasets}
\label{TraningTestSet}

The network is trained on a training set using a supervised learning method.
Training dataset consists of input vector pairs that represent the motion of each vertex through all frames and the corresponding output vector of labels which determines whether a vertex is influenced by a specific bone. The input vector size is $(3 \cdot F)$, where 3 represents the $x,y,z$ coordinates of a vertex and $F$ the number of frames for the specific animation  and the output consists of $B_{max}$ labels, where $B_{max}$ is the maximum number of bones. The current network model is fed with the training dataset and produces a result, which is then compared with the label vector, for each input vector in the training set. Based on the result of the comparison and the specific learning algorithm being used, the parameters of model are adjusted (supervised learning). 

We have used two types of training datasets, one that consists of human character animations and one consisting of animal character animations. The animal dataset contains 32 animated animal characters with an average number of 12k vertices each, an average number of 3 animations per character and an average number of 195 frames per animation.  The human dataset contains 35 animated human characters with an average number of 10k vertices each, 1 animation per character, and an average number of 158 frames per animation.

\begin{figure}[!htb]
	\centering
	\includegraphics[width=8cm]{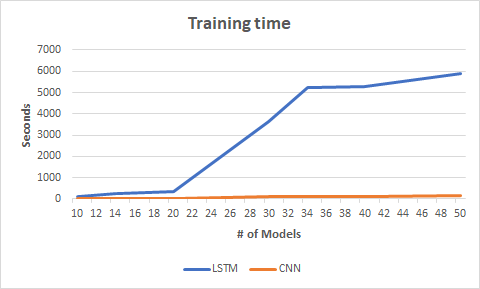}
	\caption{Training time for LSTM \& CNN networks (Section \ref{subsec:networks}).}
	\label{fig:trainingtime}
\end{figure}

Successively, the fitted network model is used to predict the response of observations in a second smaller dataset called the validation set. This set provides an unbiased evaluation of the model and has been used to tune the hyper-parameters of the network. 
Figure ~\ref{fig:trainingtime} indicates the average time that we need for training using LSTM or CNN networks (Section \ref{subsec:networks}). After a complete training session, we export the trained network model so that we can use it in our Temporal Deep Skinning method to predict bones and weights for a given animation sequence.

For the test dataset that is used in the experiments we have used a set of human and animal models that are not included in the training datasets. The efficiency of our algorithm has been tested with more than 20 human and animal models to ensure generalization and accuracy. For example the testing dataset includes four animations Spider-man (27,030 Vertices \& 28Frames), Man-Walking (15,918 Vertices \& 32 Frames),Fox (1,728 Vertices \& 400 Frames) and Lizard (29,614 Vertices \& 75 Frames) . 
Note that all dataset models are extracted from FBX animations which means that are fully animated with skeletal rigs, skinning information and transformations. The skeletal information is only used for comparison with the outcome of our method. 


\subsection{Transformation and Weight Optimization}
Approximating an animation sequence to produce a more succinct representation is common in the case of articulated models and is carried out through a process called {\em skinning}. 

For every vertex \(v_i\) that is influenced by a bone \(j\), a weight \(w_{ij}\) is assigned. For skeletal rigs the skeleton and skin of a mesh model is given in a predetermined pose also known as bind or rest pose. The rigging procedure blends the skeleton with skin which is given by the rest pose of the model. Each transformation is the concatenation of a {\em bind matrix} that takes the vertex into the local space of a given {\em bone} and a transformation matrix that moves the bone to a new position.

In LBS the new position of vertex ${v'}\,_i^{p}$ at pose (frame) $p$ is given by Equation \ref{eq:weightfit}. This approach corresponds to using proxy bones instead of the traditional hierarchical bone structure on rigid or even on deformable models \cite{10.1145/1230100.1230109}. 

\begin{equation}
\centering
{v'}\,_{i}^{p} = \sum_{j=1}^{B} w_{ij}\cdot T_{j}^{p} \cdot v_i 
\label{eq:weightfit}
\end{equation}

In this equation, \(v_i\) represents the position of the vertex in rest pose, \(w_{ij}\) the weight by which bone $j$  influences vertex $v_i$ and $T_{j}^{p}$ is the transformation that is applied to bone $j$ during frame $p$. 

\begin{figure}[H]
	\centering
	\includegraphics[width=8cm]{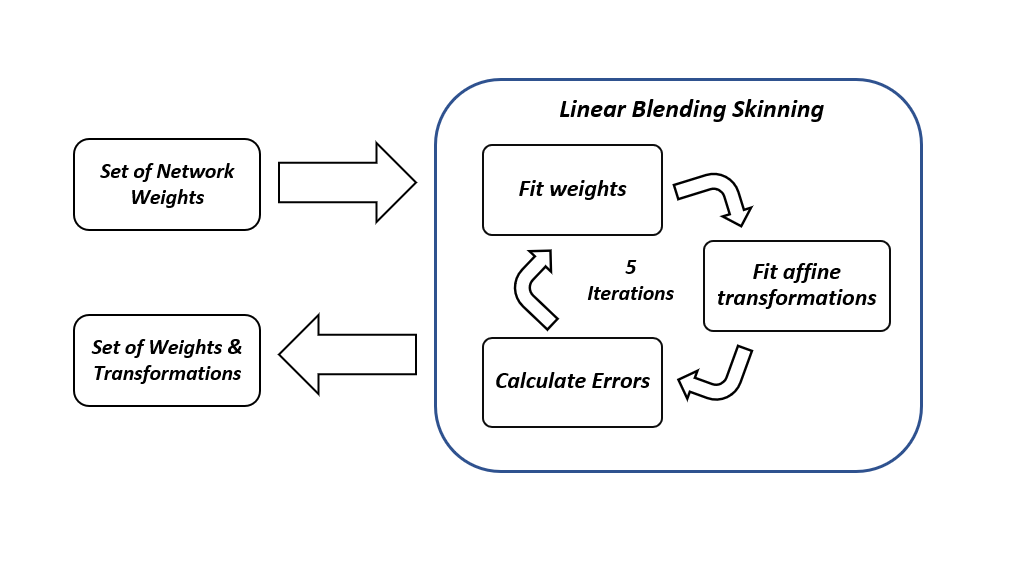}
	\caption{Deep Skinning optimization workflow for weights and transformations.}
	\label{fig:lbslabel}
\end{figure}

Figure \ref{fig:lbslabel} summarizes the successive weight and transformation optimization that aims at reducing the approximation error for all frames. 

Computing a good initial set of weights is a key step for the final result. 
In temporal deep skinning, a neural network provides the proxy bones and initial weights that are appropriate for an animation sequence. After that, we perform a successive optimization to find weights and proxy bone transformations. Both problems are formulated as least squares optimization problems that minimize the quantity given in Equation \ref{eq:transfit}. 

\begin{equation}
\centering
\sum_{i=1}^{N} {\|{v'}\,_{i}^{p} - v_{i}^{p}\|}^2 
\label{eq:transfit}
\end{equation}

where \(v_{i}^{p}\) denotes the coordinates of the original vertex in pose p, \({v'}\,_{i}^{p}\) is the approximation based on deep skinning and N is the number of vertices in the model.  
For the following, the number of vertices is $N$, the number of frames is $P$ and the number of proxy bones is $B$. To solve the weight optimization problem, we formulate the system $A\textbf{x}=b$, where matrix $A$ is a $3PN \times 6N$ (where $6$ is 6 the maximum bone number) matrix constructed by combining the rest-pose vertex positions and the corresponding transformations, $\textbf{x}$ is a $6N$ vector of unknowns that contains the weights and $b$ is a known $3PN$ vector that consists of the original (target) vertex coordinates in all frames. In the case of finding the optimal weights we include the convexity coefficient requirement as an extra equation per vertex (so  $A$ becomes $(3P+1)N \times 6N$ and $b$ becomes $(3P+1)N$). 

Finally, to solve the transformation optimization problem we formulate a linear system that consists of 3N equations, the unknowns of which are the (3x4) elements of the transformation matrices $T^p_j$ of each bone $j$ and frame $p$. This sums to $12BP$ unknowns. The system can be expressed as $A \textbf{x} = b$, where $A$ is a $3N \times 12BP$ known matrix constructed by combining the rest-pose vertex positions and the corresponding vertex weights. Moreover $b$ is a known $3N$ vector that contains the original (target) vertex coordinates. 

To avoid reverting into non linear solvers we alternatively optimize weights and transformations separately. In terms of optimization \cite{Vasilakis:2016:PPS:2949035.2949049} uses NNLS (non negative least square) optimization for enforcing the convexity condition of the weights. We express the convexity by a separate equation per vertex which is closer to the approach adopted by \cite{10.1111:j.1467-8659.2009.01602.x}. 
\cite{Vasilakis:2016:PPS:2949035.2949049} suggests that 5 iterations are enough, whereas \cite{10.1111:j.1467-8659.2009.01602.x} employs 15 iterations. We have performed experiments for up to 50 iterations and our conclusion is that after 5 iterations there is no significant error improvement. To perform the optimization problem we have employed conjugate gradient optimization which works better on multidimensional variable spaces and can be carried out efficiently on the GPU.

\subsection{Measuring the Error}
\label{subsec:errors}
We used three different types of measures to calculate error of the approximation methods. The first two measures are standard measures used in \cite{10.1145/1230100.1230109}, \cite{10.1145/1186822.1073206} and \cite{10.1111:j.1467-8659.2009.01602.x}. The first error measure is the percentage of deformation known as {\em distortion percentage (DisPer)}. 

\begin{equation}
\centering
DisPer = 100 \cdot \frac{\|A_{orig} - A_{Approx}\|_F}{\|A_{orig} - A_{avg}\|_F}.
\label{eq:distortion}
\end{equation}

where $\| \cdot \|_F$ is the Frobenius matrix metric. In Equation \ref{eq:distortion} \(A_{orig}\) is a $3NP$ matrix which consists of the real vertex coordinates in all frames of the model. Similarly, \(A_{Approx}\)  has all the approximated vertex coordinates and matrix \(A_{avg}\) contains in each column, the average of the original coordinates in all frames. \cite{10.1111:j.1467-8659.2009.01602.x} replaces 100 by 1000 and divides by the surrounding sphere diameter. Sometimes this measure tends to be sensitive to the translation of the entire character, therefore we use a different measure that is invariant to translation. The {\em root mean square (ERMS)} error measure in Equation \ref{eq:erms} is an alternative way to express distortion with the difference that we use  \(\sqrt{3NP}\) in the denominator so as to obtain the average deformation per vertex and frame during the sequence. \(3NP\) is the total number of elements in the \(A_{orig}\) matrix. \cite{Le2014} uses as denominator the diameter of the bounding box multiplied by \(\sqrt{NP}\).    

\begin{equation}
\centering
ERMS= 100 \cdot \frac{\|A_{orig} - A_{Approx}\|_F}{\sqrt{3NP}}
\label{eq:erms}
\end{equation}

We introduce a novel error measure, namely the {\em max average distance (MaxAvgDist)} given by Equation \ref{eq:maxAvg}) which is a novel quality metric that reflects better the visual quality of the result. Max distance denotes the largest vertex error in every frame. So this measure represents the average of max distances over all frames. 

\begin{equation}
\centering
MaxAvgDist = \frac{1}{P}\sum_{f=1}^{P}\max_{i=1,...,N}{\|v_{orig}^{f,i} - v_{Approx}^{f,i}\|}
\label{eq:maxAvg}
\end{equation}

Finally, we introduce an additional measure that characterizes the {\em normal distortion - (NormDistort)} and is used to measure the different behavior of two animation sequences during rendering. We compute the average difference between the original and the approximated face normals by the norm of their cross product that equals to the sine of the angle between the two normal vectors. Therefore for a model with $F$ faces and $P$ frames, where $NV^{i,j}$ is the normal vector of face $j$ at frame $i$, Equation \ref{eq:normal} computes the normal distortion measure. 

\begin{equation}
\centering
NormDistort = sin^{-1}(\frac{1}{FP} \sum_{i=1}^{P}\sum_{j=1}^{F}{||NV^{i,j}_{orig} \times NV^{i,j}_{Approx}||})
\label{eq:normal}
\end{equation}


\subsection{Building and Tuning a Neural Network for Weight Prediction}
\label{subsec:networks}
\subsubsection{Network structure}
\label{subsubsec:networks-structure}
Our method adopts a supervised learning approach to leverage the power of neural networks on multiple class classification. Consequently, we utilize a neural network instead of using clustering techniques (unsupervised learning) to obtain better initial weights and bones for skinning.
We have experimented with a variety of neural network models that can efficiently be trained to detect vertex motion patterns and mesh geometry characteristics and use similarities among them for clustering vertices into bones and determining weights implicitly through the influence of bones on the mesh surface. We have chosen networks that perform well in sequence learning. 

\begin{figure}[htb!]
	\centering
	\includegraphics[width=8cm]{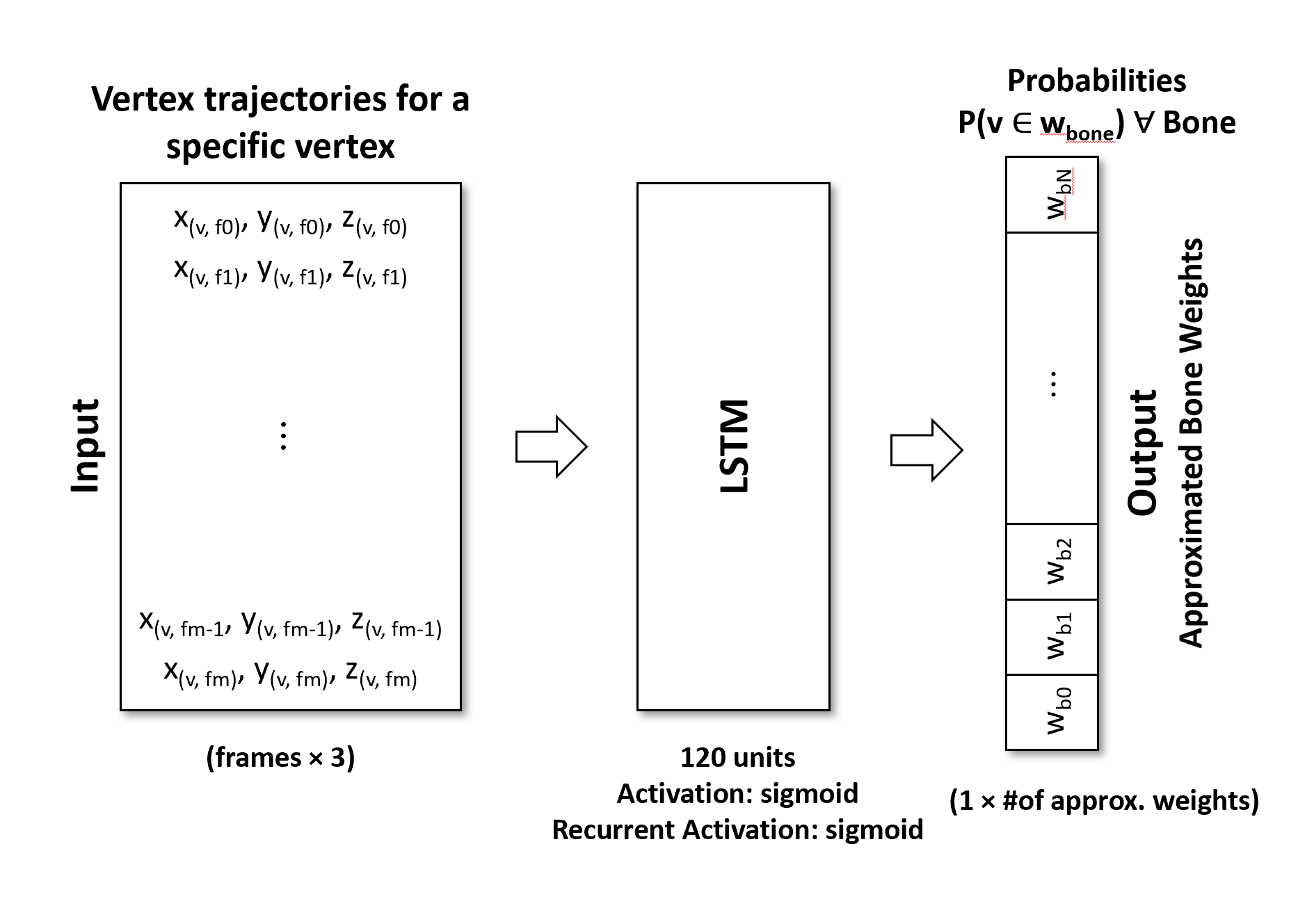}
	\caption{LSTM-Network.}
	\label{fig:LSTM}
\end{figure}


The first network that we propose as the first step and mean of animation compression is a Recurrent Neural Network (RNN). 
These networks are created by applying the same set of weights recursively over a differentiable graph-like structure by traversing it in topological order. This makes them suitable for classifying and predicting sequence data.

The type of RNN network used  is a Long Short-term Memory network firstly introduced by \cite{lstm-1997} (LSTM), which consists of units made up of a cell remembering time inconstant data values, a corresponding forget cell, an input and an output gate being responsible of controlling the flow of data in and out of the remembering component of it Figure ~\ref{fig:LSTM}. 
The actual difference of an LSTM compared to RNN is that LSTM has the capability of memorizing long-term dependencies regarding time data without resulting in emerging gradient vanishing problems (gradient loss exponentially decay). Not only does this capability make LSTM networks suitable for animation frame learning, but it is also a powerful way of predicting highly accurate weights. 

Thus, utilization of many network units for LSTM construction (120 units used) produces a network that is able of predicting weights even for models with a large number of bones. Regarding the activation functions we used (i) an alternative for the activation function (cell and hidden state) by using $sigmoid$ instead of $tanh$ and (ii) the default for the recurrent activation function (for input, forget and output gate) which is $sigmoid$. The main reason of using the $sigmoid$ function instead of the hyperbolic tangent is that our training procedure involves the network deciding per vertex whether it belongs or not to the influence range of a bone. This results in higher efficacy and additionally makes our model learn more effectively.

\begin{figure}[!htb]
	\centering
	\includegraphics[width=8cm]{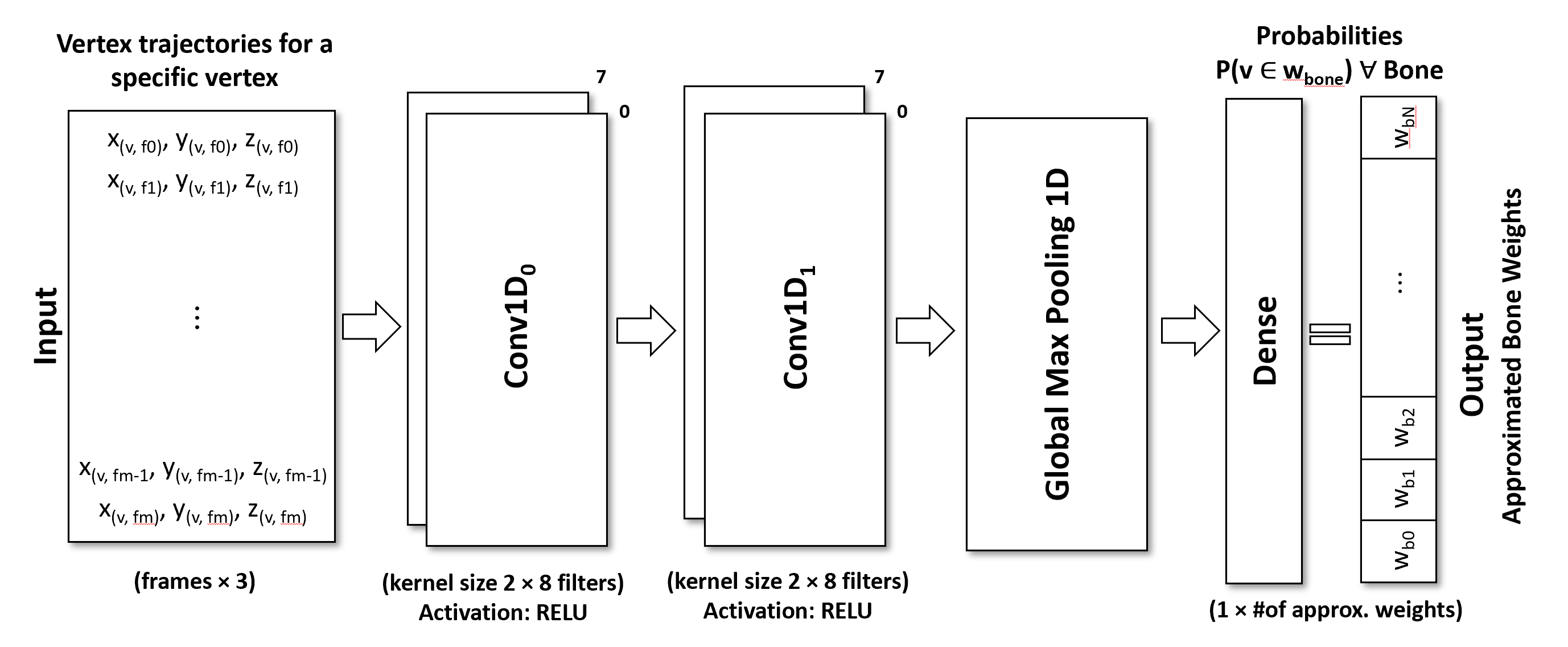}
	\caption{CNN-Network.}
	\label{fig:CNN}
\end{figure}

The second network that we have used successfully is a feed-forward network called Convolutional Neural Network (CNN) \cite{khan2019survey} that uses convolutional operations to capture patterns in order to determine classes mainly in image classification problems. CNNs are additionally able to be used in classification of sequence data with quite impressive results. On top of the two convolutional layers utilized, we have also introduced a global max-pooling layer (down-sampling layer) and a simple dense layer so that we have the desirable number of weights for each proxy bone, as it is illustrated in Figure ~\ref{fig:CNN}. In the two convolutional layers (Conv1D) used we utilize 8 filters of kernel size 2. The number of filters and kernel size have been determined experimentally. 

\begin{figure}[htb!]
	\centering
	\includegraphics[width=6cm]{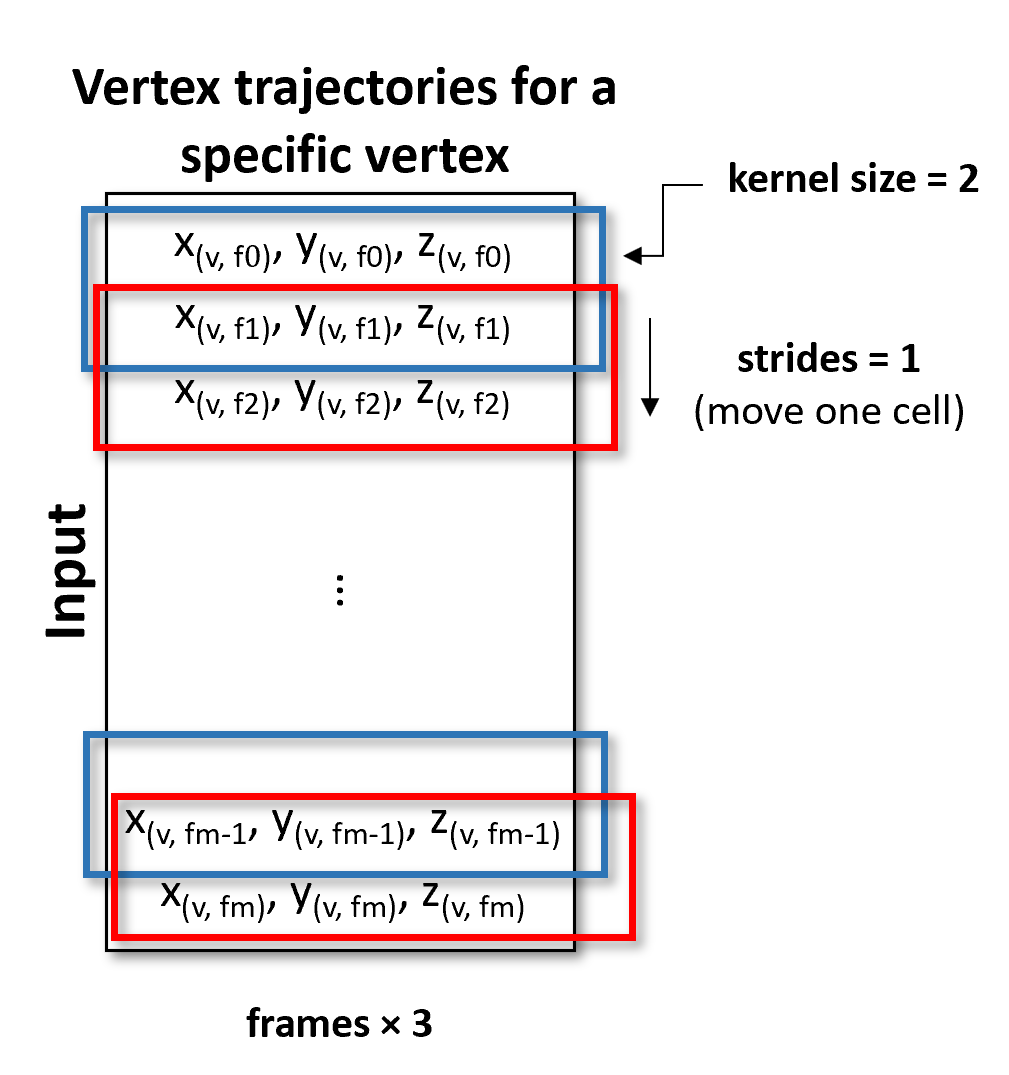}
	\caption{Convolutional kernel \& strides representation given an animation sequence input. Blue is used to highlight the previous step of computations (convolutions of input data with filter) and red the next step. In this manner, a Conv1D layer is capable of capturing vertex trajectories in an animation sequence.}
	\label{fig:CNN-kernel-strides}
\end{figure}

\begin{figure}[htb!]
	\centering
	\includegraphics[width=8cm]{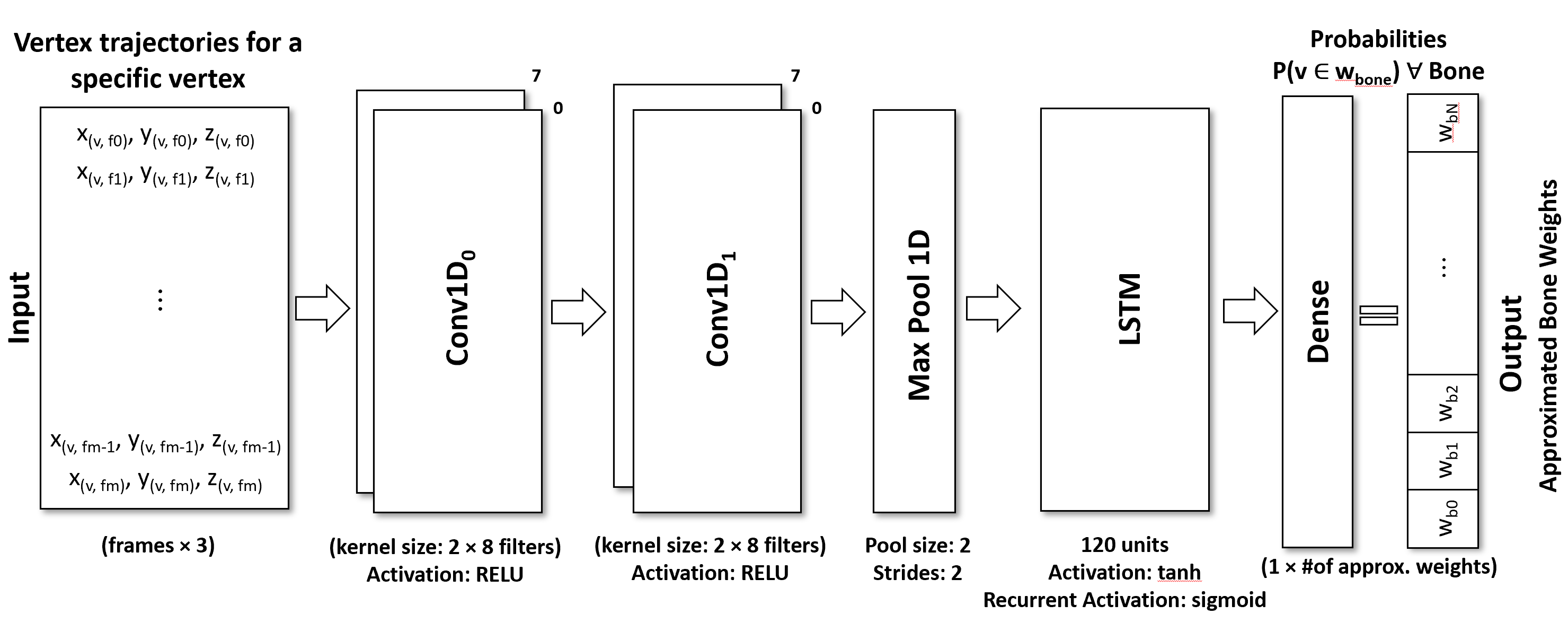}
	\caption{Hybrid-Network.}
	\label{fig:Hybrid}
\end{figure}

The last network that we have considered for completeness is a hybrid neural network (Figure ~\ref{fig:Hybrid}) that is a combination of the two aforementioned networks with some modifications. Unfortunately, the hybrid network does not perform equally well as its counterparts but it still derives comparable results.

All networks take as input an arbitrarily large sequence that represents the trajectory of a vertex, i.e. the $(x,y,z)$ position at each frame, and predict the bone weights for this vertex. 

\subsubsection{Hyper parameters tuning}
\label{subsubsec:hyperparameters-tuning}

The most essential parameters during training are (i) the effectiveness and efficiency trade off of the batch-size and (ii) the smallest number of epochs that yields maximum accuracy and minimum loss and error. 

We have determined the batch-size in a two-fold manner. 
Firstly, we have used a validation test of the vertices ($20\%$ of the examples of the training set) in all frames so as to monitor the accuracy and loss of each network model during and after training. Secondly, we have determined the best batch-size based on the skinning error of a validation dataset that consists of additional examples that do not belong to the training set.

\begin{figure}[htpb!]
	\centering
	\begin{subfigure}{0.45\columnwidth}
		\centering
		\includegraphics[width=\linewidth]{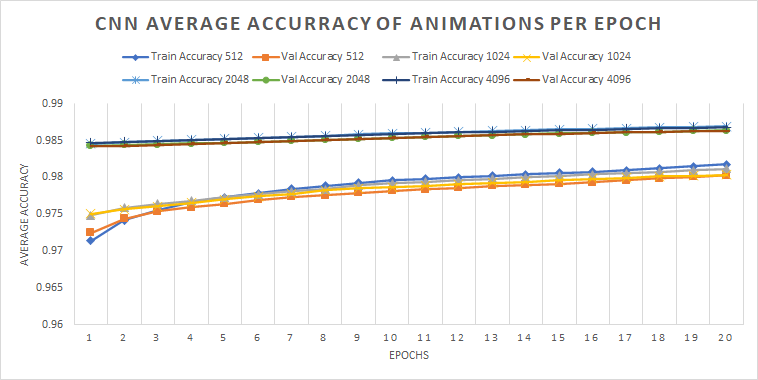}
		\caption{Binary accuracy values.}
		\label{fig:CNN-Acc}
	\end{subfigure}
	\hfill
	\begin{subfigure}{0.45\columnwidth}
		\centering
		\includegraphics[width=\linewidth]{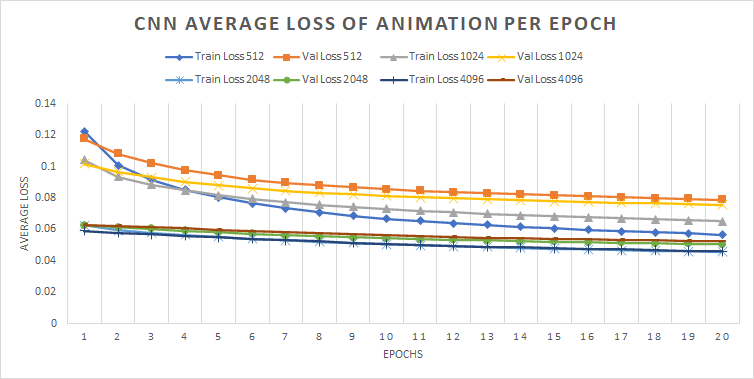}
		\caption{Binary cross-entropy loss values.}
		\label{fig:CNN-loss}
	\end{subfigure}
	\caption{Average per epoch Accuracy \& Loss for CNN.}
	\label{fig:CNN-Acc-Loss}
\end{figure}

\begin{figure}[htpb!]
	\centering
	\begin{subfigure}[htb!]{0.45\linewidth}
		\centering
		\includegraphics[width=\linewidth]{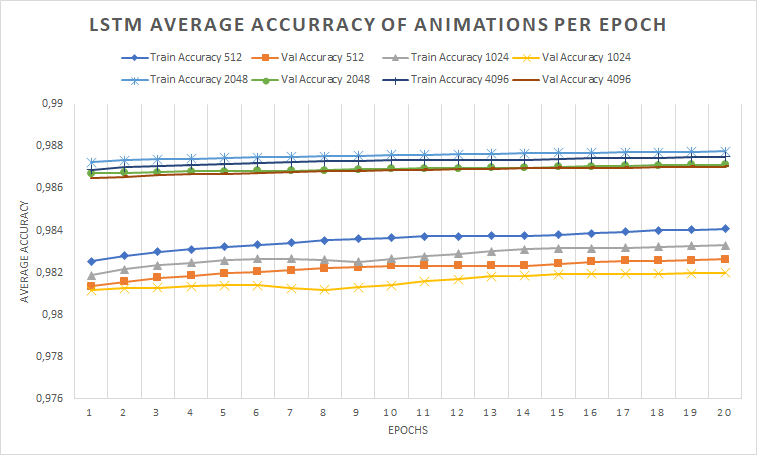}
		\caption{Binary Accuracy values.}
		\label{fig:lstm-Acc}
	\end{subfigure}
	\hfill
	\begin{subfigure}[htb!]{0.45\linewidth}
		\centering
		\includegraphics[width=\linewidth]{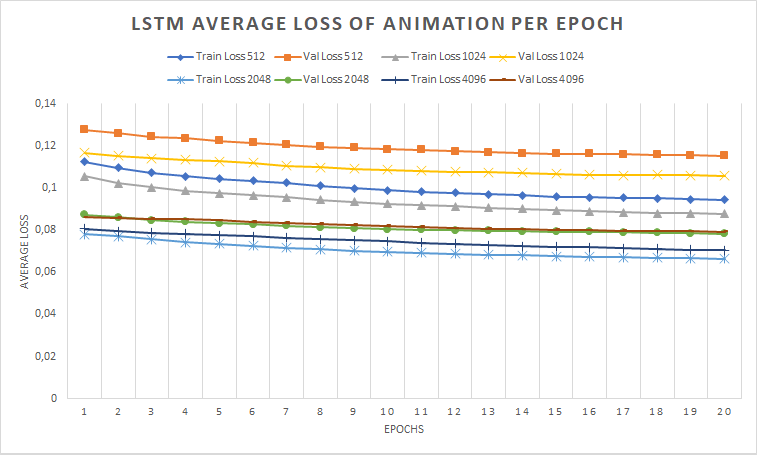}
		\caption{Binary cross-entropy loss values.}
		\label{fig:lstm-loss}
	\end{subfigure}
	\caption{Average per epoch Accuracy \& Loss for LSTM.}
	\label{fig:LSTM-Acc-Loss}
\end{figure}

Figures \ref{fig:CNN-Acc-Loss} and \ref{fig:LSTM-Acc-Loss} illustrate the loss and accuracy values that our network models achieved with several batch sizes. For loss we utilized the binary cross-entropy function given by Equation \ref{eq:binaryCrossEntropy}, since we have a multi-label problem (a vertex may belong to multiple bones).

\begin{equation}
\centering
L(y, y_{pred}) = -\frac{1}{N}\sum_{i=0}^{N} ( (1-y) \cdot log( 1-y_{pred} ) + y\cdot log(y_{pred}) )
\label{eq:binaryCrossEntropy}
\end{equation}

Where $y$ are the real values (1: belongs to a bone or 0: does not) and $y_{pred}$ are the predicted values. Binary cross-entropy measures how far in average a prediction is from the real value for every class. To this end, we also used binary accuracy which calculates the percentage of matched prediction-label pairs the 0/1 threshold value set to $0.5$. What we have inferred by these plots is that for CNN there is no reason to increase the batch-size higher than 4096 owing to the fact that accuracy and loss values tend to be almost identical after increasing batch-size from 2048 samples to 4096. Likewise, for the LSTM case (see Figure \ref{fig:LSTM-Acc-Loss}) we observe that batch-size 2048 is the best option.

From Figures \ref{fig:CNN-Acc-Loss} and \ref{fig:LSTM-Acc-Loss} we infer that we should use at least 20 epochs for training. After that the improvement of loss and accuracy is negligible but as we observed occasional overfitting is alleviated by increasing further the number of epochs.  

\begin{figure}[htpb!]
	\centering
	\begin{subfigure}[htb!]{0.45\linewidth}
		\centering
		\includegraphics[width=\linewidth]{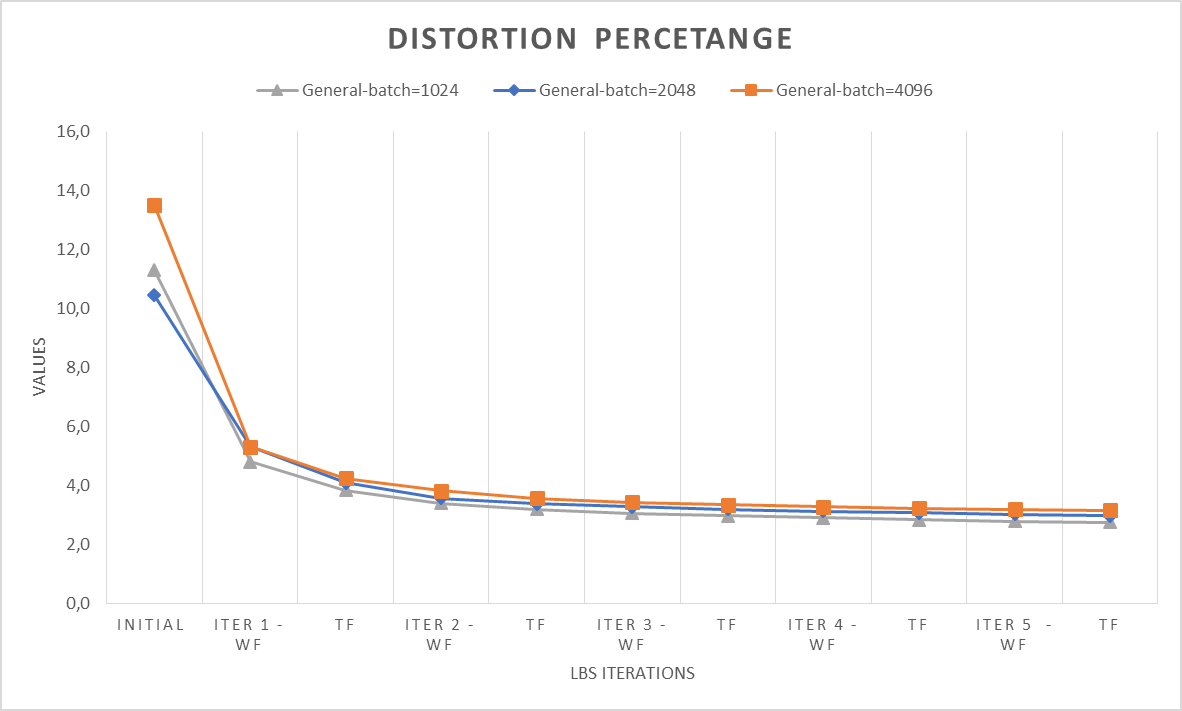}
		\caption{Distortion Percentage metric.}
		\label{fig:GeneraldisperCNN}
	\end{subfigure}
	\hfill
	\begin{subfigure}[htb!]{0.45\linewidth}
		\centering
		\includegraphics[width=\linewidth]{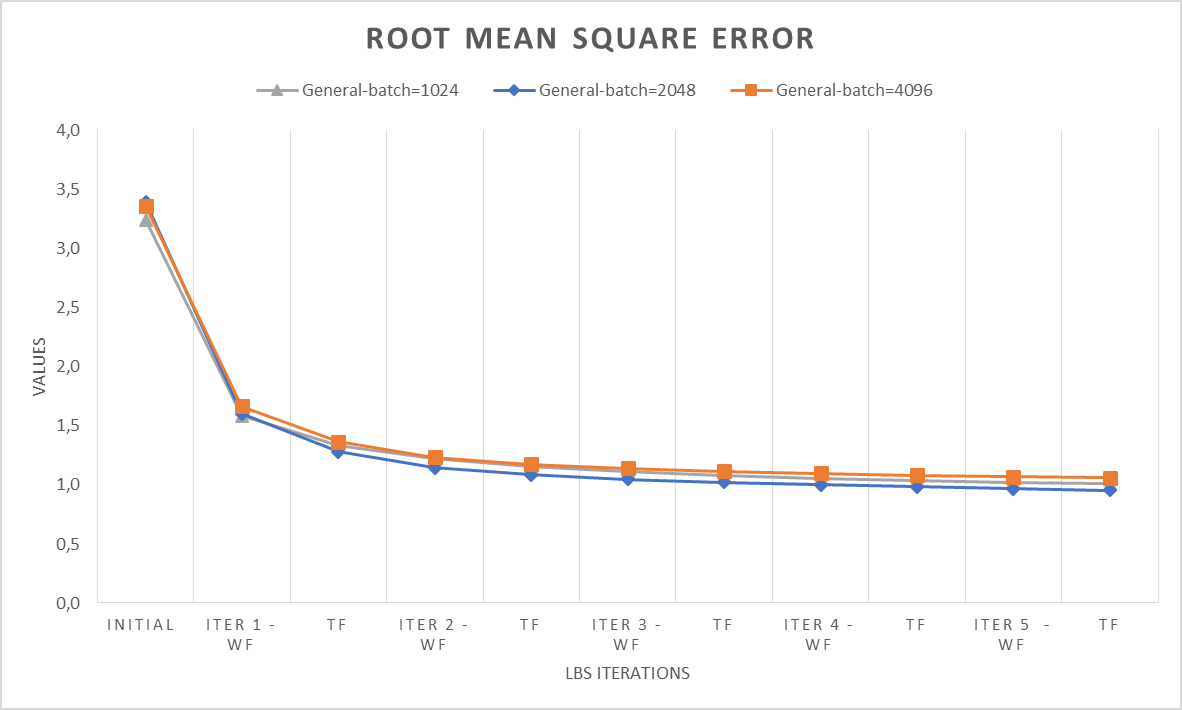}
		\caption{Root Mean Square Error metric.}
		\label{fig:GeneralrmseCNN}
	\end{subfigure}
	\hfill
	\begin{subfigure}[htb!]{0.45\linewidth}
		\centering
		\includegraphics[width=\linewidth]{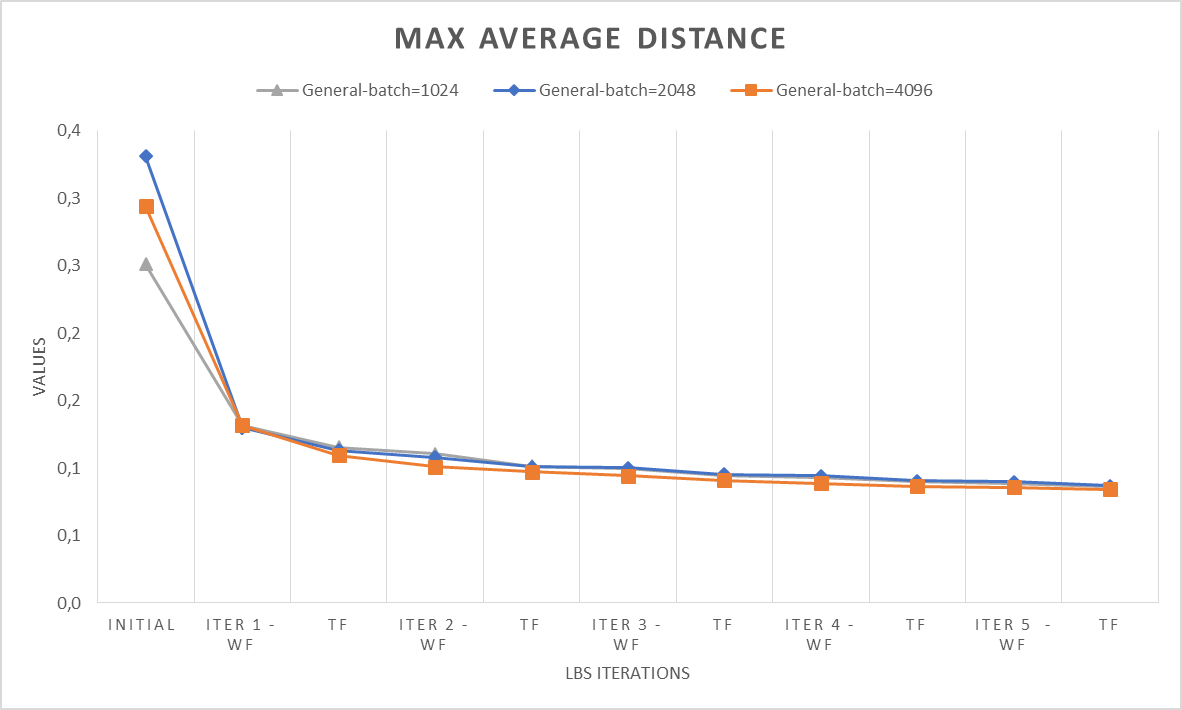}
		\caption{Max Average Distance metric.}
		\label{fig:GeneralmaxavgdisCNN}
	\end{subfigure}
	\caption{Error Metrics for batch-size tuning in CNN.}
	\label{fig:CNN-errors}
\end{figure}

\begin{figure}[htpb!]
	\centering
	\begin{subfigure}[htb!]{0.45\linewidth}
		\centering
		\includegraphics[width=\linewidth]{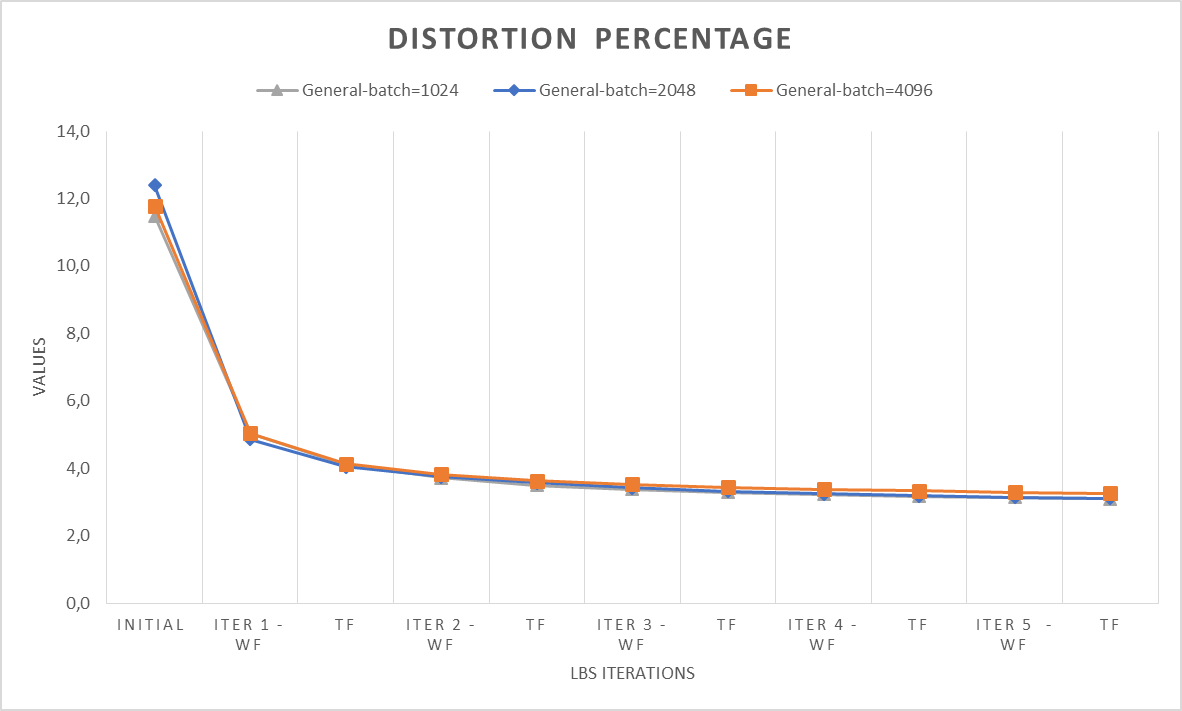}
		\caption{Distortion Percentage metric.}
		\label{fig:GeneraldisperLSTM}
	\end{subfigure}
	\hfill
	\begin{subfigure}[htb!]{0.45\linewidth}
		\centering
		\includegraphics[width=\linewidth]{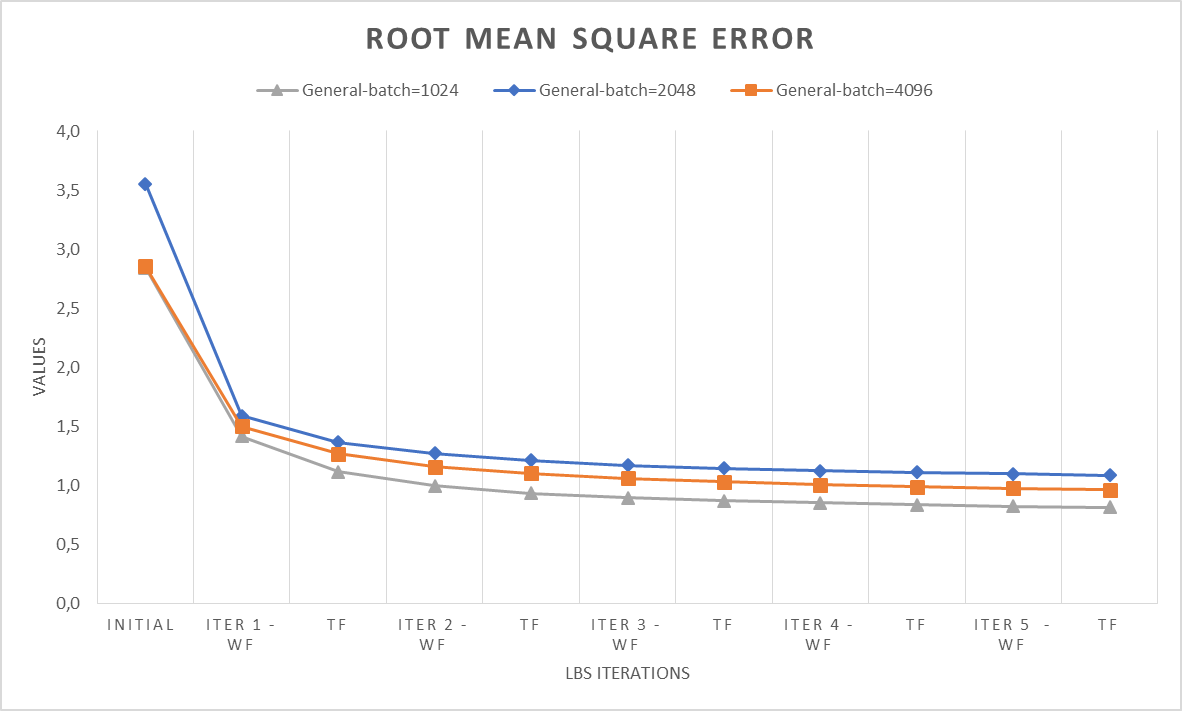}
		\caption{Root Mean Square Error metric.}
		\label{fig:GeneralrmseLSTM}
	\end{subfigure}
	\hfill
	\begin{subfigure}[htb!]{0.45\linewidth}
		\centering
		\includegraphics[width=\linewidth]{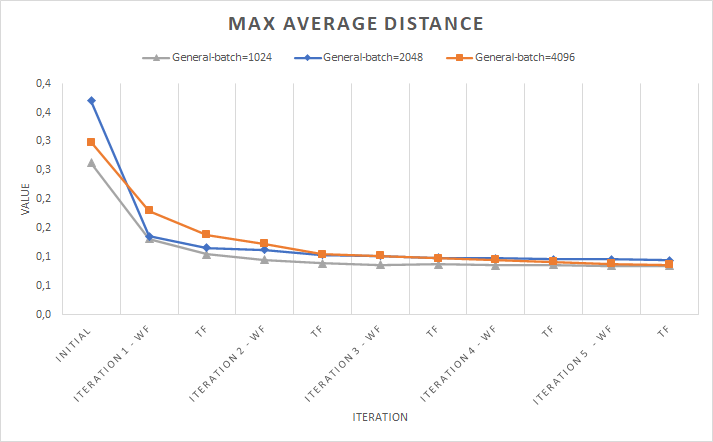}
		\caption{Max Average Distance metric.}
		\label{fig:GeneralmaxavgdisLSTM}
	\end{subfigure}
	\caption{Error Metrics for batch-size tuning in LSTM.}
	\label{fig:LSTM-errors}
\end{figure}

Furthermore, for the LSTM case, from Figure \ref{fig:LSTM-errors} we conclude that the network exhibits similar behavior to the CNN Figure \ref{fig:CNN-errors}. Therefore we select a batch-size of 4096 samples for training our LSTM model as well. 

%% file: Contents/Experiments.tex
\section{Experimental Evaluation of Deep Skinning}
One of the main contributions of our work is that it expresses a combinatorial optimization problem with constraints as a classification problem and then proposes a method to solve it using deep learning techniques. For that reason, we have conducted a thorough experimental study to substantiate the effectiveness of our method based on the resulting error.

The entire method was developed\footnote{source code available anonymously here: https://anonymous.4open.science/r/ce165fc1-66eb-4a55-8f97-267e037853d1/} using Python and Tensorflow under the Blender 2.79b scripting API. The training part runs on a system with an NVIDIA GeForce RTX 2080Ti GPU with 11GB GDDR6 RAM. We trained our network models with Adam Optimizer \cite{kingma2014adam}, $learning Rate=0.001$ for $20-100$ $epochs$ with $batchSize=4096$ over a training data-set that incorporates $60$  animated character models of different size in terms of number of vertices, animations and frames per animation. We have inferred that $20$ $epochs$ are usually enough to have our method converging in terms of the error metrics and most importantly towards an acceptable visual outcome. However to obtain better RMS and distortion errors without over-fitting $100$ $epochs$ is a safe choice independently of the training set size. Furthermore, with this choice of batch-size we overcome the over-fitting problem that was apparent by observing the Max Average Distance metric and was manifested by locally distorted meshes.

The rest of our algorithm (prediction and optimization) was developed and ran on a commodity computer equipped with an Intel Core i7-4930K 3.4GHz processor with 48Gb under Windows 10 64-bit operating System. In addition, the FESAM algorithm was developed and ran on the same system. 

\subsection{Quantitative Results}
In this section we present quantitative results for the Temporal Deep Skinning algorithm. We have conducted several experiments with multiple neural network structures to derive the top three choices of classification networks that fit best our training data with generalization capability. 

\begin{figure}[htpb!]
	\centering
	\begin{subfigure}[htb!]{0.45\linewidth}
		\centering
		\includegraphics[width=\linewidth]{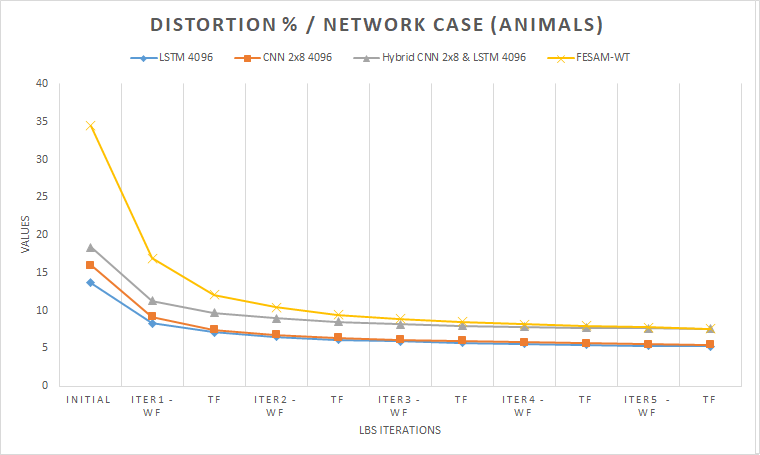}
		\caption{Distortion percentage error metric for animal characters.}
		\label{fig:distPerAnimals}
	\end{subfigure}
	\hfill
	\begin{subfigure}[htb!]{0.45\linewidth}
		\centering
		\includegraphics[width=\linewidth]{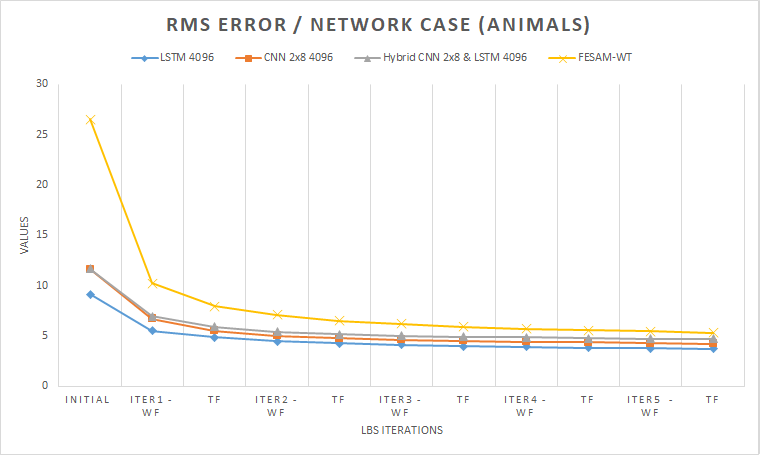}
		\caption{Root mean square error metric for animals.}
		\label{fig:rmsAnimal}
	\end{subfigure}
	\caption{Quantitative error results for animal characters.}
	\label{fig:QuantitativeAnimals}
\end{figure}

\begin{figure}[h]
	\centering
	\begin{subfigure}[htb!]{0.45\linewidth}
		\centering
		\includegraphics[width=\linewidth]{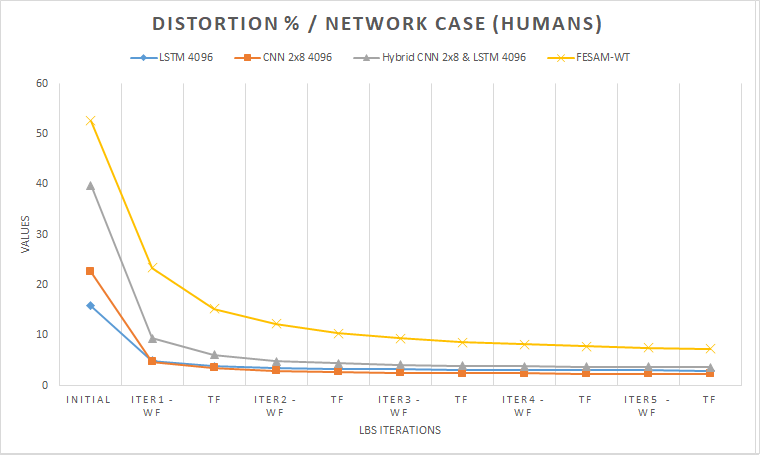}
		\caption{Distortion percentage error metric on human characters.}
		\label{fig:distPerHumans}
	\end{subfigure}
	\hfill
	\begin{subfigure}[htb!]{0.45\linewidth}
		\centering
		\includegraphics[width=\linewidth]{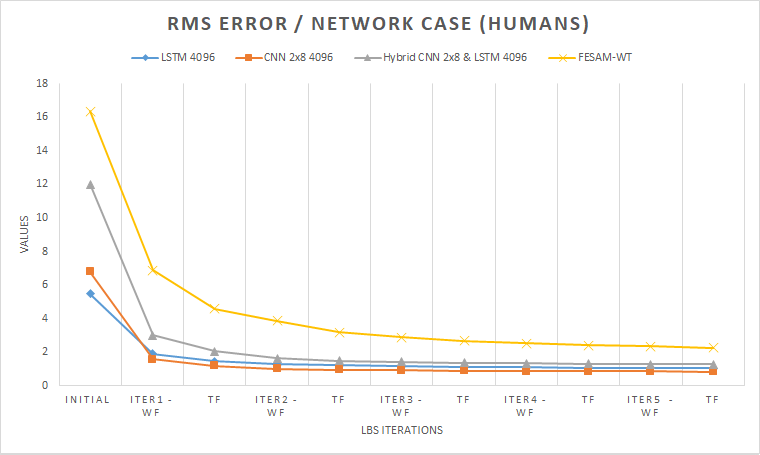}
		\caption{Root mean square error metric on human characters.}
		\label{fig:rmsHumans}
	\end{subfigure}
	\caption{Quantitative error results for human characters.} 
	\label{fig:QuantitativeHumans}
\end{figure}

The fitting part of our method optimizes weights and transformations (Weight Fitting-WF \& Transformation Fitting-TF in figures \ref{fig:CNN-errors}, \ref{fig:LSTM-errors}, \ref{fig:QuantitativeAnimals}, \ref{fig:QuantitativeHumans}, \ref{fig:NormResults}) alternatively using a linear optimizer for five iterations. After five iterations we have observed that there is practically no improvement of the error for any of the methods. For each iteration the error metrics are computed and registered for every fitting case separately. Additionally, the initial error values in the plots below are the actual errors computed with the weights that each neural network produces.

Figures \ref{fig:QuantitativeAnimals} and \ref{fig:QuantitativeHumans} provide a comparative evaluation of our method with the most competent skinning approach FESAM \cite{10.1111:j.1467-8659.2009.01602.x}. The original FESAM algorithm follows three steps of optimization with the first step being the process of optimizing the initial pose something that is not compatible with the traditional animation pipelines. For that reason this step is not included in our experiments and subsequently we use the FESAM-WT approach with two steps (weight and transformation optimization). More specifically, we evaluate the performance of our approach with distortion and RMS errors for the three top networks as compared to FESAM-WT. Based on Figure \ref{fig:distPerAnimals} and Figure \ref{fig:rmsAnimal} we conclude that LSTM 4096 performs overall better among the three prevalent networks on animal characters.  We have performed the same experiments with the same test sets using decimated versions of our animation characters ($50\%$ and $20\%$ decimation) and we have obtained the same results with differences only in the fourth decimal place. 

The behavior of our method on human characters is illustrated in Figures \ref{fig:distPerHumans} and \ref{fig:rmsHumans}. We conclude that CNN 4096 is the most appropriate network structure in comparison with the other two. 

Table \ref{fig:comparisons} is a comparison of our method with other similar methods producing LBS schemes with pseudo bones when presented with several benchmark animation sequences from literature. More specifically, presents a comparison of our method on four benchmark animation sequences, that were not produced by fully animated rigs, with all previous combinations of LBS, quaternion-based and SVD methods. $N$ is the number of Vertices, $F$ is the number of frames and the number in round brackets is the result of the method combined with SVD. Our method derives better results in terms of both error and compression rate as compared to methods I-IV. Method V is only cited for reference since it only obtains compression and is not compatible with any of the standard animation pipelines.

\begin{table*}[htpb!]
	\centering
	\begin{adjustbox}{max width=\linewidth}
		\begin{tabular}{|c|c|c|c|c|c|c|c|c|c|c|c|c|c|c|c|c|}
			\hline
			\multicolumn{17}{|c|}{{\color[HTML]{333333} \textbf{Approximation Error ERMS}}}                                                                                                                                                                                                                                                                                                                                                                                                                                                              \\ \hline
			\multicolumn{3}{|c|}{{\color[HTML]{333333} \textbf{Input Data}}} & \multicolumn{2}{c|}{{\color[HTML]{333333} \textbf{Our Method}}} & \multicolumn{2}{c|}{{\color[HTML]{333333} \textbf{Method I}}} & \multicolumn{2}{c|}{{\color[HTML]{333333} \textbf{Method II}}} & \multicolumn{2}{c|}{{\color[HTML]{333333} \textbf{Method III}}} & \multicolumn{2}{c|}{{\color[HTML]{333333} \textbf{Method IV}}} & \multicolumn{2}{c|}{{\color[HTML]{333333} \textbf{Method V}}} & \multicolumn{2}{c|}{{\color[HTML]{333333} \textbf{Compression Rate}}} \\ \hline
			\textbf{Dataset}             & \textbf{N}     & \textbf{F}     & \textbf{Bones}                  & \textbf{ERMS}                 & \textbf{Bones}                 & \textbf{ERMS}                & \textbf{Bones}                 & \textbf{ERMS}                 & \textbf{Bones}                  & \textbf{ERMS}                 & \textbf{Bones}                 & \textbf{ERMS}                 & \textbf{Bones}                 & \textbf{ERMS}                & \textbf{OURS}                     & \textbf{I-IV}                     \\ \hline
			\textbf{Horse-gallop}        & 8,431           & 48             & 26                              & 0.15                          & 30                             & 2.3(0.3)                     & 30                             & 4.9(2.9)                      & 30                              & 1.3                           & 30                             & 2.4                           & -                              & 2E-5                         & 92.5                              & 92.3                              \\ \hline
			\textbf{Elephant-gallop}     & 42,321          & 48             & 18                              & 0.35                          & 25                             & 2.6(0.5)                     & 25                             & 15(6.5)                       & 25                              & 1.4                           & 25                             & 2.3                           & -                              & 6E-5                         & 93.59                             & 93.51                             \\ \hline
			\textbf{Camel-gallop}        & 21,887          & 48             & 16                              & 0.22                          & 23                             & 3.1(0.5)                     & 23                             & 4.7(2.2)                      & 23                              & 1.4                           & 23                             & 2.8                           & -                              & 2E-4                         & 93.45                             & 93.33                             \\ \hline
			\textbf{Samba}               & 9,971           & 175            & 17                              & 0.60                          & 30                             & 8.6(3.6)                     & 30                             & 11.4(6)                       & 30                              & 1.5                           & 30                             & 4                             & -                              & 0.2                          & 97.6                              & 97.4                              \\ \hline
		\end{tabular}
	\end{adjustbox}
	\caption{Comparative evaluation of our method versus Method I \cite{10.1145/1186822.1073206}, Method II \cite{10.1145/1230100.1230109}, Method III \cite{10.1111:j.1467-8659.2009.01602.x}, Method IV \cite{Sattler2005}, Method V \cite{Alexa2000}.}
	\label{fig:comparisons}
\end{table*}

\begin{figure}[htpb!]
	\centering
	\includegraphics[width=6cm]{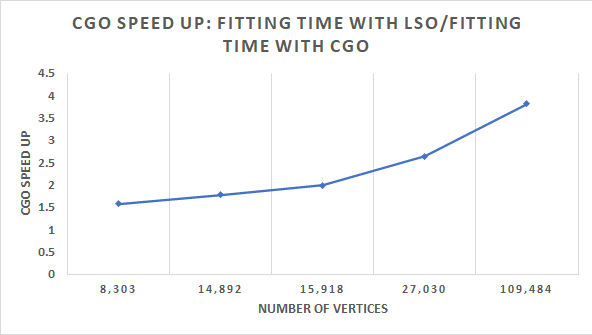}
	\caption{Speed up of fitting time by using the conjugate gradient optimization method.}
	\label{fig:speedup}
\end{figure}

Finally, Figure \ref{fig:speedup} shows the speed up that we have achieved in the fitting time by using the conjugate gradient method which is more efficient in multi dimensional problems such as the ones that we are solving ($12BP$ variables for transformations, and $6N$ variables for weights).

\subsection{Visual Quality Evaluation Results}

In computer graphics qualitative results (visual and otherwise) is an important means of assessing a novel method. In this section we present three processes for assessing the visual quality of temporal deep skinning.

\subsubsection{Quality measure evaluation}
We use the MaxAvgDist quality assessment measure that indicates how far in terms of visual quality the generated frames are from the original frame sequence (see Section \ref{subsec:errors}). Low measure values correspond to high quality animation. 

\begin{figure}[htpb!]
	\centering
	\begin{subfigure}[htb!]{0.45\linewidth}
		\centering
		\includegraphics[width=\linewidth]{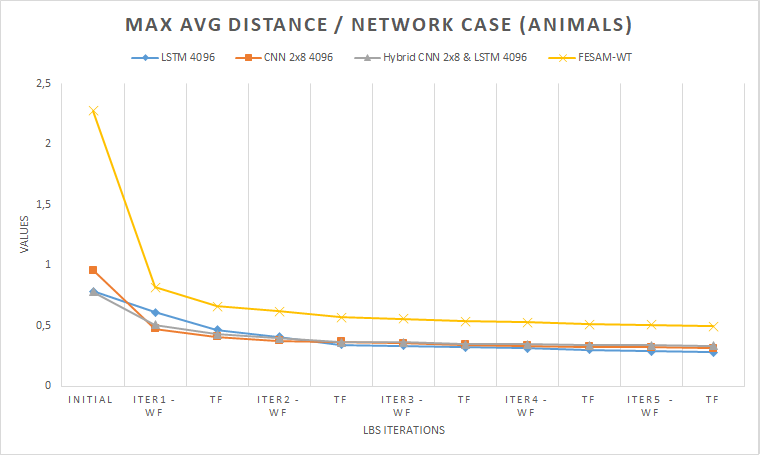}
		\caption{Max average distance on animal characters.}
		\label{fig:MaxAnimals}
	\end{subfigure}
	\hfill
	\begin{subfigure}[htb!]{0.45\linewidth}
		\centering
		\includegraphics[width=\linewidth]{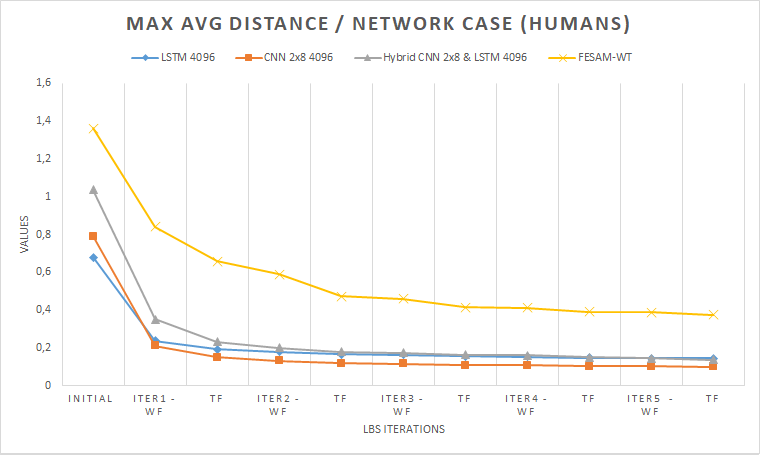}
		\caption{Max average distance on human characters.}
		\label{fig:Maxhumans}
	\end{subfigure}
	\caption{Qualitative error metric $MaxAvgDistr$ results for humans and animals.}
	\label{fig:Qualitative1}
\end{figure}

Figure \ref{fig:Qualitative1} suggests that Temporal Deep Skinning yields results with better quality measure as compared to FESAM-WT. The results of Figures \ref{fig:MaxAnimals} and \ref{fig:Maxhumans} confirm the quantitative results. Specifically, the LSTM network for animals and the CNN for humans are the most appropriate choices quantitatively and qualitatively.

\subsubsection{Visualization-based Evaluation}

As an additional assessment criterion for our method we provide an illustration of the visual outcome. By using the term visual outcome, we refer to the approximated output frames as compared to the frames of original 3D model. After conducting several experiments we have observed that our approach seems to approximate better the original model. In every case there is a noticeable difference between temporal deep skinning and FESAM-WT. To this end, a demonstration video is also provided as supplementary material.

\begin{figure}[htpb!]
	\centering
	\begin{subfigure}[htb!]{0.7\linewidth}
		\centering
		\includegraphics[width=\linewidth]{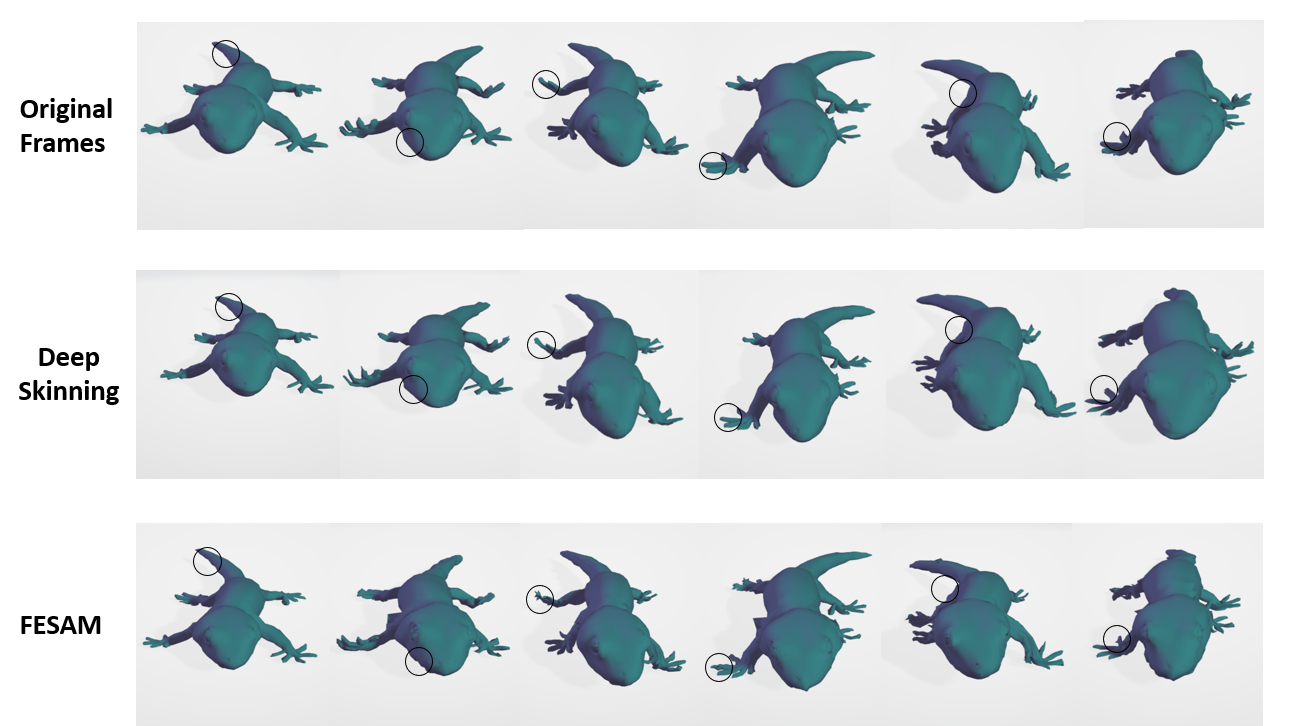}
		\label{fig:lizard}
	\end{subfigure}
	\begin{subfigure}[h]{0.7\linewidth}
		\centering
		\includegraphics[width=\linewidth]{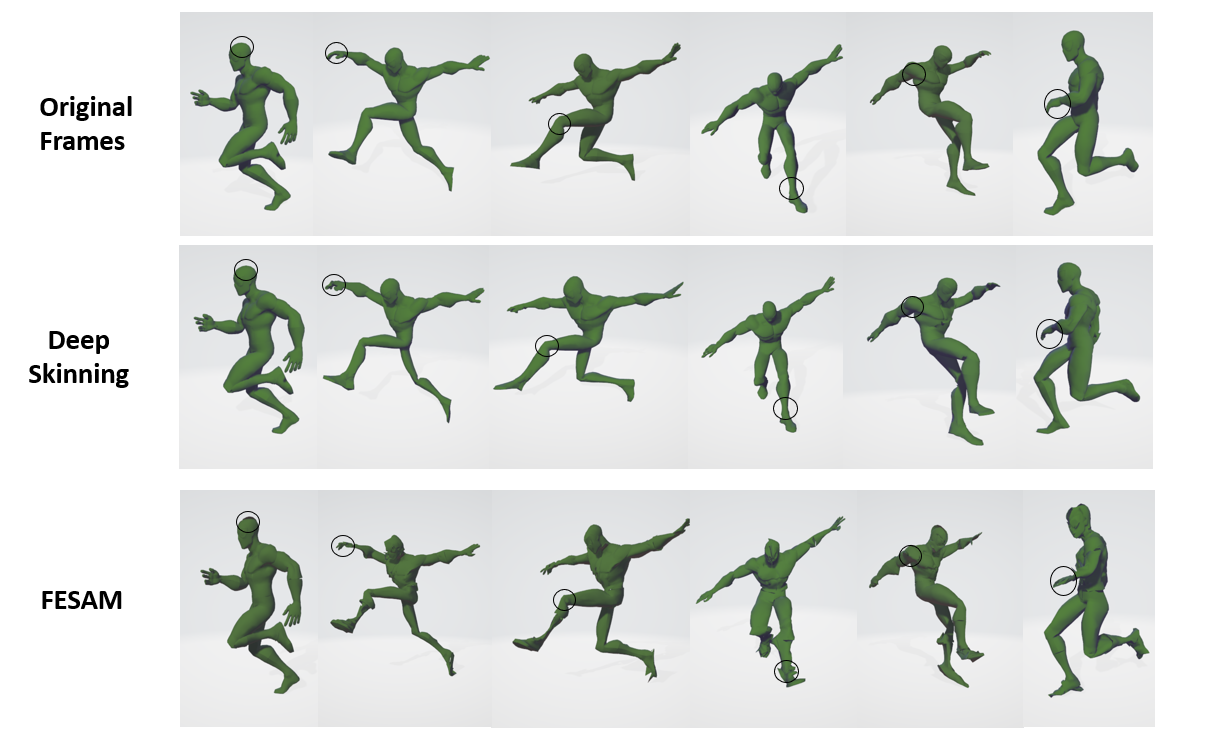}
		\label{fig:spideman}
	\end{subfigure}
	\caption{Visual comparison of  Deep Skinning, FESAM-WT and the original frames for two models.Six frames have been selected in which structural flaws are marked by small circles.}
	\label{fig:Qualitative}
\end{figure}

Figure ~\ref{fig:Qualitative} illustrates the differences of the two approximation methods as compared to the original model animation. Several frames have been selected with noticeable structural flaws.      

\begin{figure}[htpb!]
	\centering
	\includegraphics[width=8cm]{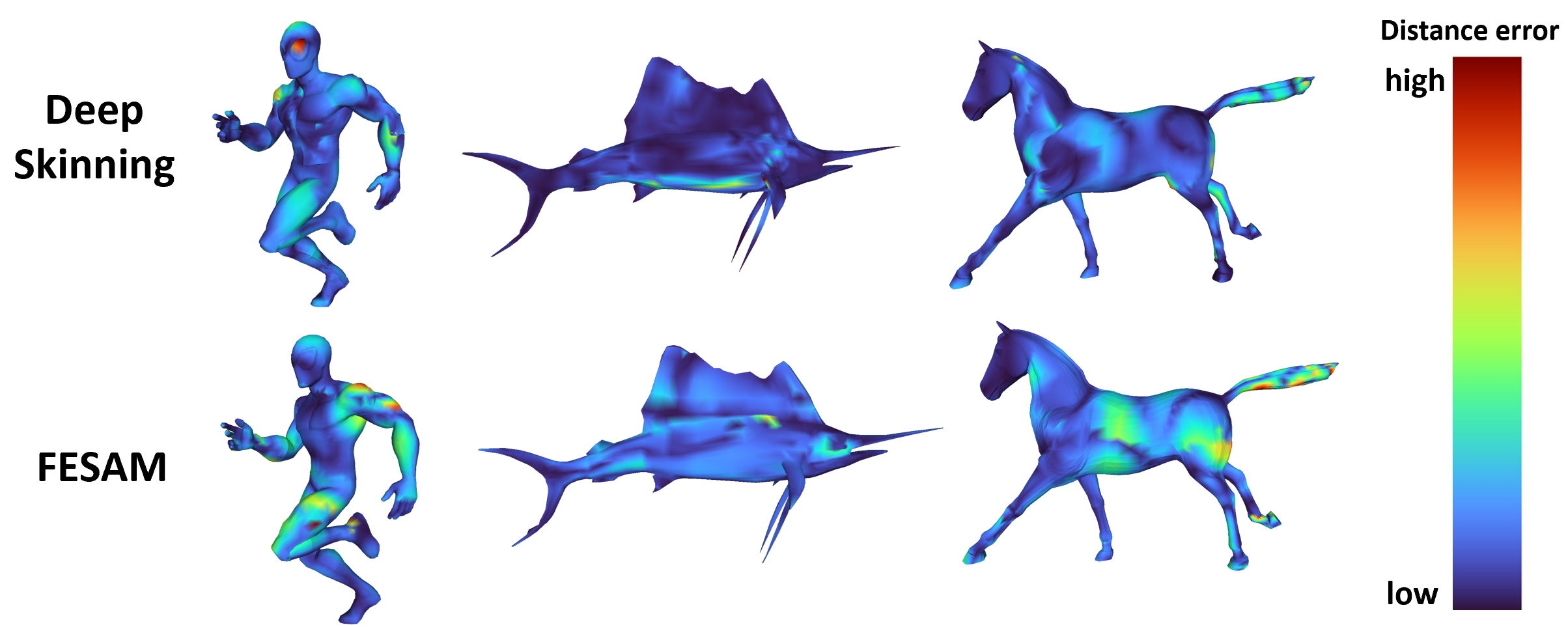}
	\caption{Distance error comparison in a particular frame between Deep Skinning and FESAM-WT.}
	\label{fig:viualComp}
\end{figure}

Error visualization techniques can provide an insight for the parts where errors occur.
We use the turbo colormap\cite{turbo-colormap} to represent the error per vertex which is color blind friendly. This error is the distance in a particular frame of the approximated vertex from the original one. Figure \ref{fig:viualComp} illustrates the per vertex error in a particular frame for deep skinning and FESAM\_WT.  

\subsubsection{Lighting Quality Evaluation}

\begin{figure}[htpb!]
	\centering
	\begin{subfigure}{0.45\linewidth}
		\centering
		\includegraphics[width=\linewidth]{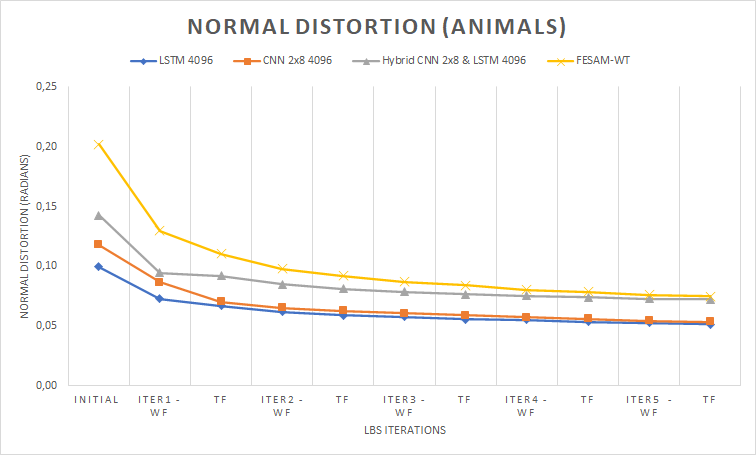}
		\caption{Normal distortion on animal characters.}
		\label{fig:NormAnimals}
	\end{subfigure}
	\begin{subfigure}{0.45\linewidth}
		\centering
		\includegraphics[width=\linewidth]{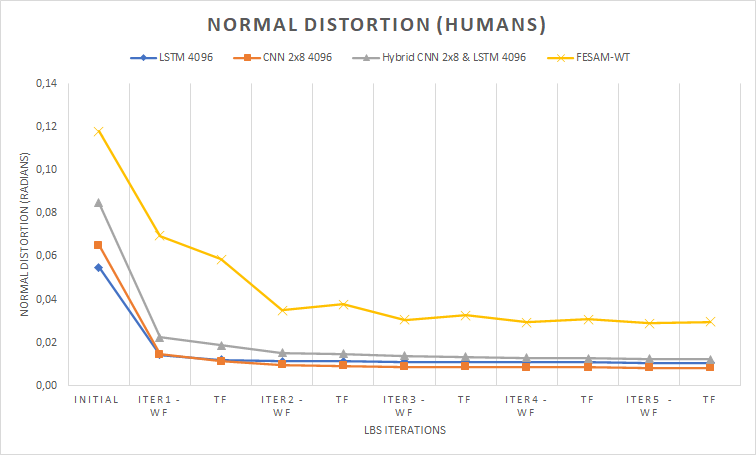}
		\caption{Normal distortion on human characters.}
		\label{fig:NormHumans}
	\end{subfigure}
	\caption{Qualitative normal distortion metric results for humans and animals.}
	\label{fig:NormResults}
\end{figure}

Finally, we offer the results of evaluating the average distortion of normal vectors. The normal distortion measure (see section \ref{subsec:errors}) shows how close the normal vectors of the approximated sequence are to the normal vectors of the original animation sequence. This determines how the approximated character will behave in an lighting model as compared to the original animated character. The results of Figure \ref{fig:NormResults} exhibit an average error of $0.01$ radians for human characters and an average error of $0.05$ radians for animal characters.

\subsection{Discussion and Applications}

We have presented a method called {\em Temporal Deep Skinning}  that feeds an animation sequence with no underlying skeleton or rigging/skinning information to a pre-trained neural network to generate an approximated compressed skinning model with pseudo-bones. 

\begin{table*}[htpb!]
	\centering
	\begin{adjustbox}{max width=\textwidth}
		\begin{tabular}{|c|c|c|c|c|c|c|c|c|c|c|c|c|c|c|c|c|c|}
			\hline
			\multicolumn{3}{|c|}{{\color[HTML]{333333} \textit{\textbf{Input Data}}}} & \multicolumn{15}{c|}{{\color[HTML]{333333} \textit{\textbf{Approximation Error ERMS}}}}                                                                                                                                                                                                                                                                                      \\ \hline
			\multicolumn{3}{|c|}{{\color[HTML]{333333} \textit{\textbf{}}}}            & \multicolumn{3}{c|}{{\color[HTML]{333333} \textit{\textbf{Our Method}}}} & \multicolumn{3}{c|}{{\color[HTML]{333333} \textit{\textbf{Method A}}}} & \multicolumn{3}{c|}{{\color[HTML]{333333} \textit{\textbf{Method B}}}} & \multicolumn{3}{c|}{{\color[HTML]{333333} \textit{\textbf{Method C}}}} & \multicolumn{3}{c|}{{\color[HTML]{333333} \textit{\textbf{Method D}}}} \\ \hline
			\textbf{Dataset}                     & \textbf{N}       & \textbf{F}       & \textbf{Bones}          & \textbf{ERMS}          & \textbf{CRP}         & \textbf{Bones}         & \textbf{ERMS}         & \textbf{CRP}         & \textbf{Bones}         & \textbf{ERMS}         & \textbf{CRP}         & \textbf{Bones}         & \textbf{ERMS}         & \textbf{CRP}         & \textbf{Bones}         & \textbf{ERMS}         & \textbf{CRP}         \\ \hline
			\textit{\textbf{Horse-gallop}}       & 8,431             & 48               & 26                      & 0.15                   & 92.5                  & 27                     & 0.19                  & 92.4                  & 27                     & 0.44                  & 92.4                  & 27                     & 1.10                  & 92.4                  & 27                     & 0.88                  & 92.4                  \\ \hline
			\textit{\textbf{Samba}}              & 9,971             & 175              & 17                      & 0.60                   & 97.6                  & 22                     & 0.63                  & 97.4                  & 22                     & 1.29                  & 97.4                  & 22                     & 1.57                  & 97.4                  & 22                     & 1.79                  & 97.4                  \\ \hline
		\end{tabular}
	\end{adjustbox}
	\caption{Comparison between Temporal Deep Skinning and four methods. Specifically Method A \cite{Le2014}, Method B \cite{SchaeferYuksel2007}, Method C \cite{Aguiar2008}, Method D \cite{Hasler2010}.}
	\label{fig:skeletonTable}
\end{table*}

\begin{figure}[htpb!]
	\centering
	\begin{subfigure}{0.45\linewidth}
		\centering
		\includegraphics[width=\linewidth]{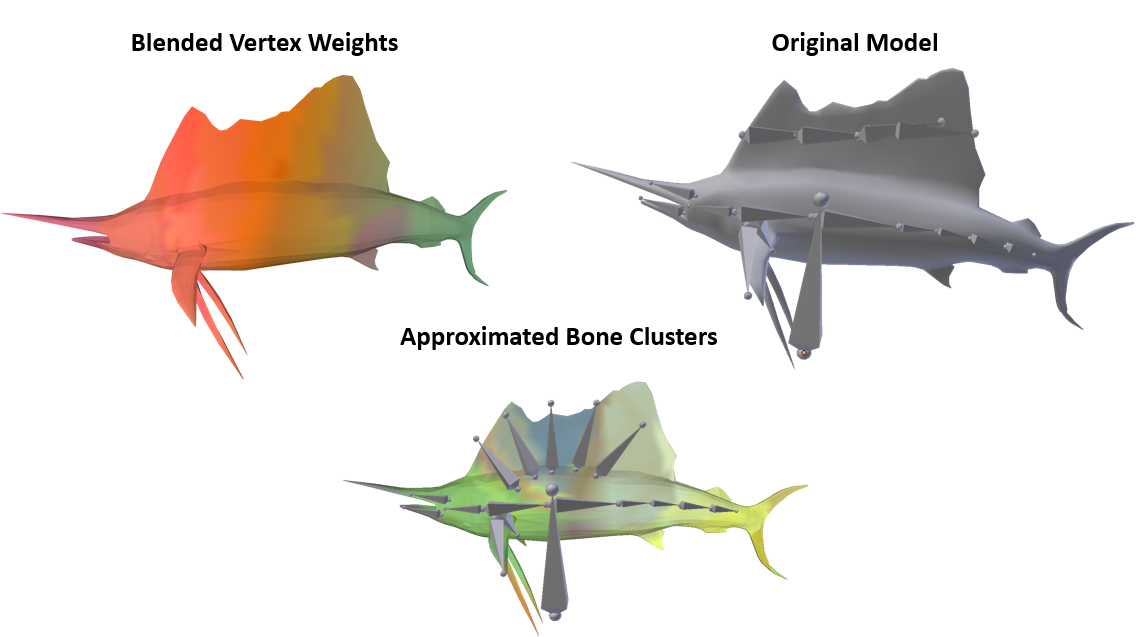}
		\caption{Original and approximate representation for a Sailfish.}
		\label{fig:sailfish}
	\end{subfigure}
	\begin{subfigure}{0.45\linewidth}
		\centering
		\includegraphics[width=\linewidth]{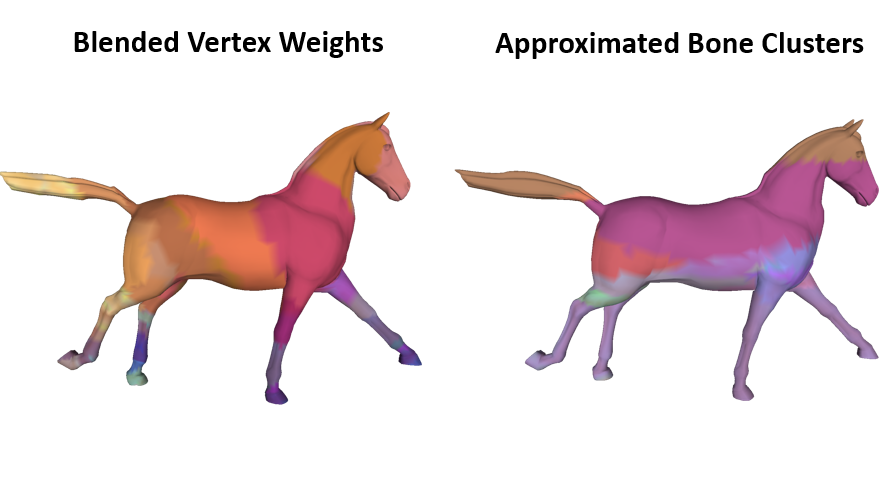}
		\caption{Approximate bone clusters and vertex weights for an animation sequence.}
		\label{fig:horse}
	\end{subfigure}
	\label{fig:horse2}
	\caption{Original and approximate representations.}
\end{figure}

Moreover, we have developed a post processing tool that using the compressed skinning model with pseudo bones and per frame transformations obtained by temporal deep skinning produces the corresponding hierarchical skeleton, skinning data and transformations. More specifically, using the mesh clustering derived by our method, the pseudo bones and transformations we produce a fully animated character model. This is accomplished by the following steps: (i) perform weight regularization and derive disjoint vertex clusters that are influenced by each bone, (ii) based on the neighboring clusters export the joints of the entire model (the structure of the skeleton) \cite{10.1145/2366145.2366218} and (iii) finally perform joint location adjustment by geometric constraints and a simple recalculation of orientation and rotation for each of the joints that yields their final position \cite{AGR16}. Figure \ref{fig:sailfish} illustrates the original and approximated representation of a 3D model. This animated model consisting of $14,007$ vertices and $4,669$ faces was approximated by the deep skinning algorithm with $24$ bone clusters and up to six weights per vertex. 

Figure \ref{fig:horse} presents the computed bones and weights for an animation sequence. This animation sequence consists of 48 different frames from the horse-gallop sequence. After the Deep Skinning algorithm we were able to extract 19 bone clusters and up to six weights per vertex. Subsequently, we have produced a fully animated character.

Table \ref{fig:skeletonTable} provides a comparison of our method with four methods that produce actual skeletal rigs. In this case of \ref{fig:skeletonTable} we cite the results from the papers since such methods are difficult to reproduce and this goes beyond the scope of this paper. For two models ({\em horse gallop} and {\em samba}) we have measured the $ERMS$ error and the compression rate percentage (CRP).  Note that the results of \cite{Le2014} were converted to our ERMS metric by multiplying by $\frac{D}{\sqrt{3}}$, where $D$ is the diagonal of the bounding box of the rest pose.

For the horse gallop model our method approximates the sequence by using 26 bones and achieves a smaller $ERMS$ error as compared to all previous methods. For the samba model our method uses 17 bones and outperforms all previous methods.

%% file: Contents/Conclusions.tex
\section{Conclusions}
We have introduced  a novel approach that derives pseudo-bones and weights for an animated sequence using deep learning
on a training set of fully rigged animated characters. 
We have experimented with a variety of neural network models that can efficiently be trained to detect vertex motion patterns and mesh geometry characteristics and exploit similarities among them for clustering vertices into bones and determining weights implicitly through the influence of bones on the mesh surface. 
Our method does not require setting or tuning any parameters regarding the mesh structure or the kinematics of the animation. 

Based on a comparative evaluation, we conclude that the approximation error of our method is always smaller than the error of previous approaches that are compatible with existing animation pipelines.
